\def\Gmm{{\bf\Gamma}}
\def\gmm{\mbox{\boldmath$\gamma$}}
\def\eqn#1{\eqno(\rm#1)}
\def\tH{\tilde{\bf H}}
\def\ht{\tilde{\bf h}}
\def\mi{\bigskip\noindent}
\def\al{ {\it et al}.~}
\def\la{\left\langle}
\def\ra{\right\rangle}
\def\lb{\left\{}
\def\rb{\right\}}
\def\rf#1{(\ref{#1})}
\documentstyle[12pt,psfig]{article}
\begin{document}
{\center
{\it Short title: LARGE-SCALE MAGNETIC FIELD GENERATION}

\mi{\bf GENERATION OF MULTISCALE MAGNETIC FIELD

BY PARITY-INVARIANT TIME-PERIODIC FLOWS}

\mi
V.A. Zheligovsky$^{a,b,c,}$\footnote{E-mail: vlad@mitp.ru},
O.M. Podvigina$^{a,b,c,}$\footnote{E-mail: olgap@mitp.ru}

\mi$^a$Observatoire de la C\^ote d'Azur, CNRS U.M.R. 6529,\\
BP 4229, 06304 Nice Cedex 4, France

\mi$^b$International Institute of Earthquake Prediction Theory\\
and Mathematical Geophysics,\\
79 bldg.2, Warshavskoe ave., 117556 Moscow, Russian Federation

\mi$^c$Laboratory of general aerodynamics, Institute of Mechanics,\\
Lomonosov Moscow State University,\\
1, Michurinsky ave., 119899 Moscow, Russian Federation

\bigskip
Submitted to {\it Geophysical and Astrophysical Fluid Dynamics}

}

\bigskip{\bf Abstract.} We study generation of magnetic fields involving large
spatial scales by time- and space-periodic small-scale parity-invariant flows.
The anisotropic magnetic eddy diffusivity tensor is calculated by the standard procedure
involving expansion of magnetic modes and their growth rates in power series
in the scale ratio. Our simulations, conducted for flows with random harmonic
composition and exponentially decaying energy spectra, demonstrate that
enlargement of the spatial scale of magnetic field is
beneficial for generation by time-periodic flows. However,
they turn out, in general, to be less efficient dynamos, than steady flows.


\vspace*{5cm}

\pagebreak
The present work is an extension of the studies carried out by
Lanotte\al(2000) and Zheligovsky\al(2001), who found steady parity-invariant
flows with a negative magnetic eddy diffusivity to be quite common.
A similar investigation for turbulent flows is desirable,
especially given that flows in experimental dynamos are necessarily turbulent
(see discussion {\it ibid.}). Some time-dependent flows are known to be unable
to sustain negative magnetic eddy diffusivity -- namely, the flows,
$\delta$-correlated in time at each point in space. They advect
a mean magnetic field as a passive scalar, and therefore eddy diffusivity can
only exceed molecular diffusivity (Biferale\al, 1995).
Thus the question, whether time-dependent flows
can give rise to negative magnetic eddy diffusivity, is not trivial.

However, dynamo simulations for time-dependent turbulent flows are numerically
demanding. For this reason we focus our attention here at an ``intermediate"
class of flows -- those periodic in time, and employ the simplest of them:
\begin{equation}
{\bf v}({\bf x},t)={\bf U}({\bf x})+\sqrt{\omega}
({\bf V}_c({\bf x})\cos\omega t+{\bf V}_s({\bf x})\sin\omega t).
\label{realflow}\end{equation}
Like in the cited papers, a flow is supposed to be
$2\pi$-periodic in spatial variables $\bf x$ and parity-invariant, i.e.
\begin{equation}
{\bf v}({\bf x},t)=-{\bf v}(-{\bf x},t).
\label{parity}\end{equation}
(Consequently, its space-averaged helicity is zero,
and no $\alpha$-effect is present.)

Two families of space- and time-periodic flows were closely examined
in the context of fast kinematic magnetic dynamo theory. Both are
generalisations of ABC flows, and flows from both families are constructed
of a small number of spatial Fourier harmonics. A flow of the kind of ``modulated
waves" was considered by Otani (1993) and Childress \& Gilbert (1995); it
belongs to the class \rf{realflow}. ``Circularly polarised" flows
were employed in simulations of Galloway \& Proctor (1992) and Galloway \&
O'Brian (1993); near-integrable flows of this kind were studied analytically
by Ponty\al (1993, 1995) and Childress \& Gilbert (1995).
The assumed time dependence was responsible for chaotic behaviour of the flow
trajectories (this is necessary for a dynamo to be fast), though the flows
depended only on two spatial variables (for steady flows this
rules out chaos). This gave an opportunity to
separate out dependence on the third spatial coordinate and
to make convincing computations for high magnetic Reynolds numbers $R_m\sim 10^4$
(Galloway \& Proctor, 1992; Galloway \& O'Brian, 1993).
Circularly polarised non-integrable ABC flows depending on three spatial
variables were considered by Brummell\al(1999, 2001).

We consider here a class of flows which seem to be more physically realistic:
like Zheligovsky\al(2001), we employ flows with a random spatial harmonic
composition and an exponentially decaying energy
spectrum. (Random-harmonic flows with a slow -- hyperbolic --
energy spectrum decay were also explored {\it ibid.}, but they were found
to be less efficient generators of large-scale magnetic field; consequently,
such flows are not investigated here.) They can be regarded as a ``poor man's
model" of turbulent flows. Ideally we would like to examine dynamo properties
of a large number of sample flows from the class under consideration, so that
to collect statistically sound information. Unfortunately, computations
even in the case of the simple time dependence \rf{realflow}
are numerically expensive: evaluation of one instance of magnetic eddy
diffusivity requires more than a day of a CPU of a Dec Alpha processor.
Hence it appears possible to evaluate only a limited number of such instances.
However, since spatial Fourier components of the flows employed in
computations are chosen at random,
one can hope that results of computations presented here are
typical and attributable to a large variety of flows.
(Note, that flows from the selected class seize to be typical if
${\bf V}_c=0$ or ${\bf V}_s=0$: it is shown in Section 3 that
a non-zero contribution from the time-periodic part of \rf{realflow}
in the limit of high frequencies requires linear independence
of the vector fields ${\bf V}_c$ and ${\bf V}_s$. Thus \rf{realflow}
represents the simplest class of time-periodic flows with hopefully a typical
behaviour.)

Also due to numerical complexity of the problem, we have performed computations
for just one value of molecular magnetic diffusivity, $\eta=0.1$~.
The following constraints restrict the choice of $\eta$: On the one hand,
if it is too large, diffusion dominates, inhibiting generation of magnetic
field. On the other, if molecular diffusion is smaller than the threshold
for the onset of generation of a small-scale magnetic field (i.e., magnetic
field which has the same spatial periodicity, as the flow $\bf v$),
the short-scale instability dominates: the large-scale instability becomes
``negligible", since growth rates of large-scale magnetic modes are
infinitesimally small (large-scale magnetic modes being perturbations of
small-scale modes with non-vanishing spatial means, associated with a zero
growth rate). Zheligovsky\al(2001) explored large-scale kinematic dynamos
for three values of molecular magnetic diffusivity, $\eta=0.1$, 0.2 and 0.3
(the flow was normalised so that its r.m.s. was 1). For all the three values
no generation of small-scale magnetic field occurred, and
for $\eta=0.1$ more flows sustained negative magnetic eddy diffusivity, than
for $\eta=0.2$ and 0.3~. This has suggested to choose the value $\eta=0.1$
in our simulations; like {\it ibid.}, we have checked
that $\eta=0.1$ is above the magnetic diffusivity threshold for the onset
of generation of a small-scale zero-mean magnetic field for every flow,
for which magnetic eddy diffusivity has been computed.

In Section 1 we present a mathematical statement of the problem and state
results of derivation of the magnetic eddy diffusivity tensor
(details can be found in Appendix).
Dependencies of the minimal magnetic diffusivity on the ratio
of energies of the steady and time-dependent parts of the flow
and on temporal frequency $\omega$ are studied numerically.
Results of simulations are discussed in Section 2.
Steady flows are found to be more capable of magnetic field
generation, than time-periodic ones.
Owing to the factor $\sqrt{\omega}$ in \rf{realflow}, in the limit of high
temporal frequencies the time-dependent part of the
flow provides a finite contribution to the magnetic eddy diffusivity tensor.
In Section 3 the limit magnetic eddy diffusivity tensor is formally derived
for $\omega\to\infty$, and it is shown numerically that flows can have negative
eddy diffusivity for high frequencies.
Our results are briefly summarized in the Conclusion.

\bigskip
\noindent
{\bf 1. Magnetic eddy diffusivity tensor for time-periodic flows}

\bigskip
In this Section we consider a kinematic dynamo problem for a flow ${\bf v}({\bf x},t)$
of time period $T$, which is $2\pi$-periodic in each Cartesian variable in space,
solenoidal ($\nabla\cdot{\bf v}=0$) and parity-invariant \rf{parity}.

Temporal evolution of a magnetic field $\bf h$ is described by
the magnetic induction equation
\begin{equation}
{\partial{\bf h}\over\partial t}=
\eta\nabla^2{\bf h}+\nabla\times({\bf v}\times{\bf h}).
\label{indeqn}\end{equation}
Substituting ${\bf h}={\bf H}({\bf x},t)e^{\lambda t}$ into \rf{indeqn}
one finds that a magnetic mode ${\bf H}({\bf x},t)$ satisfies
\begin{equation}
\lambda{\bf H}=-{\partial{\bf H}\over\partial t}+
\eta\nabla^2{\bf H}+\nabla\times({\bf v}\times{\bf H}),
\label{floquet}\end{equation}
i.e.~the kinematic dynamo problem reduces to a Floquet problem for
the magnetic induction operator. The mode ${\bf H}({\bf x},t)$ is solenoidal,
\begin{equation}
\nabla\cdot{\bf H}=0,
\label{Hsolen}\end{equation}
and it is assumed to have the same time period $T$, as the flow.
Re$\lambda$ is then the average rate of growth (or decay) of the mode in time.

We consider magnetic modes, involving large spatial scales.
The modes are supposed to depend on the fast spatial variable $\bf x$ and
on the slow variable $\bf y=\epsilon\bf x$. By the chain rule, spatial derivatives
in the eigenmode equation \rf{floquet} and the solenoidality condition \rf{Hsolen}
must be modified:
\begin{equation}
\nabla\to\nabla_{\bf x}+\epsilon\nabla_{\bf y}
\label{grad}\end{equation}
(the subscripts $\bf x$ and $\bf y$ refer
to differentiation in fast and slow variables, respectively).
The ratio of the two scales, $\epsilon>0$, is a small parameter of the problem.

Solution to the Floquet problem \rf{floquet} is sought in the form of power series
\begin{equation}
\lambda=\sum_{n=0}^\infty\lambda_n\epsilon^n,
\label{grate}\end{equation}
\begin{equation}
{\bf H}=\sum_{n=0}^\infty\tH_n({\bf x},{\bf y},t)\epsilon^n.
\label{mmode}\end{equation}
Substituting \rf{grad} and the series \rf{grate} and \rf{mmode}
into \rf{floquet}, expanding and equating coefficients for each power of $\epsilon$,
one obtains a hierarchy of equations. It is discussed in Appendix, how all terms
of \rf{grate} and \rf{mmode} can be determined by a systematic procedure.
In particular, it is shown that:

\mi
$\bullet$ The leading term in \rf{grate} is $\lambda_2$ and the expansion involves terms
with even indices only ($\lambda_0=\lambda_{2n+1}=0$ for any integer $n\ge 0$).

\mi
$\bullet$ All terms of \rf{mmode} with even indices are parity anti-invariant in
fast variables ($\tH_{2n}(-{\bf x},{\bf y},t)=\tH_{2n}({\bf x},{\bf y},t)$
for any integer $n\ge 0$), and all terms with odd indices are parity-invariant in
fast variables \hbox{($\tH_{2n+1}(-{\bf x},{\bf y},t)=-\tH_{2n+1}({\bf x},{\bf y},t)$}
for any integer $n\ge 0$).

\mi
$\bullet$ The leading term in the decomposition of an eigenmode is independent
of time: $\tH_0=\tH_0({\bf x},{\bf y})$. Its spatial average satisfies the eigenvalue
equation for the anisotropic magnetic eddy diffusivity operator,
\begin{equation}
\eta\nabla^2_{\bf y}\la\tH_0\ra+\nabla_{\bf y}\times
\sum_{k=1}^3\sum_{m=1}^3{\bf D}_{m,k}{\partial\la\tH^k_0\ra
\over\partial y_m}=\lambda_2\la\tH_0\ra,
\label{leadingtermseq}\end{equation}
and the solenoidality condition
\begin{equation}
\nabla_{\bf y}\cdot\la\tH_0\ra=0.
\label{leadingsol}\end{equation}
Here $\la\cdot\ra$ denotes the mean of a vector field over the cube of periodicity:
$$\la{\bf f}\ra\equiv(2\pi)^{-3}\int_{[0,2\pi]^3}{\bf f}({\bf x},{\bf y})d{\bf x},$$
and superscripts enumerate Cartesian components of a vector field.

Coefficients of the eddy diffusivity tensor can be evaluated using solutions
of two auxiliary problems:
\begin{equation}
-{\partial{\bf S}_k\over\partial t}
+\eta\nabla^2{\bf S}_k+\nabla\times({\bf v}\times{\bf S}_k)
+{\partial{\bf v}\over\partial x_k}={\bf 0}
\label{prob1}\end{equation}
({\em the first auxiliary problem}), and
\begin{equation}
-{\partial\Gmm_{m,k}\over\partial t}
+\eta\nabla^2\Gmm_{m,k}+\nabla\times({\bf v}\times\Gmm_{m,k})
+2\eta{\partial{\bf S}_k\over\partial x_m}
+{\bf e}_m\times({\bf v}\times({\bf S}_k+{\bf e}_k))={\bf 0}
\label{prob2}\end{equation}
({\em the second auxiliary problem}); here \{${\bf e}_m$\}
is the basis of Cartesian unit vectors. Then
$${\bf D}_{m,k}={1\over T}\int_0^T\la{\bf v}\times\Gmm_{m,k}\ra dt.$$

It can be verified that ${\bf S}_k$ are parity anti-invariant
(${\bf S}_k({\bf x},t)={\bf S}_k(-{\bf x},t)$) and
solenoidal; $\Gmm_{m,k}$ are parity-invariant and satisfy
$\nabla_{\bf x}\cdot\Gmm_{m,k}+{\bf S}^m_k=0$.

The partial differential operator in the left-hand side
of \rf{leadingtermseq} is comprised of second-order derivatives
with constant coefficients. Consequently, its eigenvectors are Fourier harmonics:
$\la\tH_0\ra=\ht e^{i\bf qy}$, where
${\bf q}$ is a (constant) wavevector, and $\ht$ satisfies
\begin{equation}
{\bf q}\times\sum_{k=1}^3\sum_{m=1}^3{\bf D}_{m,k}{\bf q}_m\ht_k
+\eta|{\bf q}|^2\ht=-\lambda_2\ht,
\label{oldeigen}\end{equation}
\begin{equation}
\ht\cdot{\bf q}=0.
\label{ortho}\end{equation}
The quantity $\eta_{\rm eddy}=\min_{|{\bf q}|=1}(-\lambda_2)$
is regarded as the minimal magnetic eddy diffusivity.
When it is negative, the associated magnetic mode grows in time.

\pagebreak
\noindent
{\bf 2. Magnetic eddy diffusivity for flows \rf{realflow}: numerical results}

\bigskip
Numerical simulations have been carried out for flows \rf{realflow},
where ${\bf U,\ V}_c$ and ${\bf V}_s$ are $2\pi$-periodic parity-invariant
solenoidal fields. The fields have been generated by the procedure which was
applied by Zheligovsky\al(2001): ($i$) a half of Fourier harmonics
with random uniformly distributed components are generated (all the rest ones
are obtained by complex conjugation, so that the vector field is real),
($ii$) the gradient part of the resultant field is projected out,
and ($iii$) the harmonics are rescaled in each Fourier spherical shell
to obtain the desirable energy spectrum. The spectrum of flows employed in our simulations
exponentially decreases by 6 orders of magnitude, the Fourier series
being truncated at wavenumber 10. The vector fields are normalised so that
\begin{equation}
E_{\rm total}=1,
\label{enorm}\end{equation}
where
$$E_{\rm total}\equiv {1\over T}\int_0^T(2\pi)^{-3}\int_{[0,2\pi]^3}|{\bf v}|^2d{\bf x}dt
=\int_{[0,2\pi]^3}\left(|{\bf U}|^2+{1\over2}(|{\bf V}_c|^2+|{\bf V}_s|^2)\right)d{\bf x}$$
is the average total energy of the flow for $\omega=1$
(hence the magnetic Reynolds number can be estimated as $R_m=\eta^{-1}$).
Equipartition of the average energy of the two time-dependent terms
in \rf{realflow} is assumed:
\begin{equation}
\int_{[0,2\pi]^3}|{\bf V}_c|^2d{\bf x}=\int_{[0,2\pi]^3}|{\bf V}_s|^2d{\bf x}.
\label{equipart}\end{equation}

Solutions to the auxiliary problems \rf{prob1} and \rf{prob2} are
sought in the form of Fourier series, with the $64^3$ Fourier harmonics resolution
in space and 8 harmonics resolution in time. For this resolution the spatial energy
spectra of the solutions decay by at least 10 orders
of magnitude, and the temporal ones -- by 4-5 orders of magnitude.

All computations presented in this paper have been performed for
molecular magnetic diffusivity $\eta=0.1$~.
It has been verified that for each sample flow \rf{realflow},
for which a magnetic eddy diffusivity value is reported here,
the real part of the dominant eigenvalue of the
magnetic induction operator acting in the space of $2\pi$-periodic small-scale
(i.e.~independent of the slow variables) zero-mean magnetic fields
is negative, i.e. $\eta=0.1$ is above the magnetic diffusivity threshold
for the onset of generation of a small-scale magnetic field.
Codes of Zheligovsky (1993a) have been applied for numerical treatment
of the Floquet problems\footnote{
Note that in the Floquet problem \rf{floquet} eigenmodes are defined
up to a factor $e^{iJ\omega t}$, where $J$ is an arbitrary integer,
the associated eigenvalues differing by $iJ\omega$. Hence in computations
precautions must be taken to converge to the dominant eigenvalue,
whose imaginary part does not exceed $\omega/2$ in absolute value,
i.e.~whose imaginary part is minimal in absolute value in this set
of equivalent eigenmodes.} for small-scale magnetic modes.

The following questions have been addressed.

\vskip 2mm
\noindent
a) How does the minimal magnetic eddy diffusivity $\eta_{\rm eddy}$ change
when the flow \rf{realflow} is close to a steady one?
Magnetic eddy diffusivity has been evaluated for
temporal frequency $\omega=1$ and molecular viscosity $\eta=0.1$ for
30 independent samples of \rf{realflow} satisfying \rf{enorm} and \rf{equipart},
such that the ratio of the average energy of the time-dependent part of \rf{realflow},
$$E_{\rm osc}={1\over2}\int_{[0,2\pi]^3}(|{\bf V}_c|^2+|{\bf V}_s|^2)d{\bf x},$$
to the energy of the steady profile is small:
$$E_{\rm osc}\left/\int_{[0,2\pi]^3}|{\bf U}|^2d{\bf x}\right.=1/400.$$
A histogram of the values of $\delta\eta_{\rm eddy}$, the amount of change in magnetic
eddy diffusivity due to introduction of this weak time periodicity into the flow,
is shown on Fig.~1. Only in 2 cases out of 30 the
moderate time dependence makes $\eta_{\rm eddy}$ to decrease.

\vskip6mm
\centerline{\psfig{file=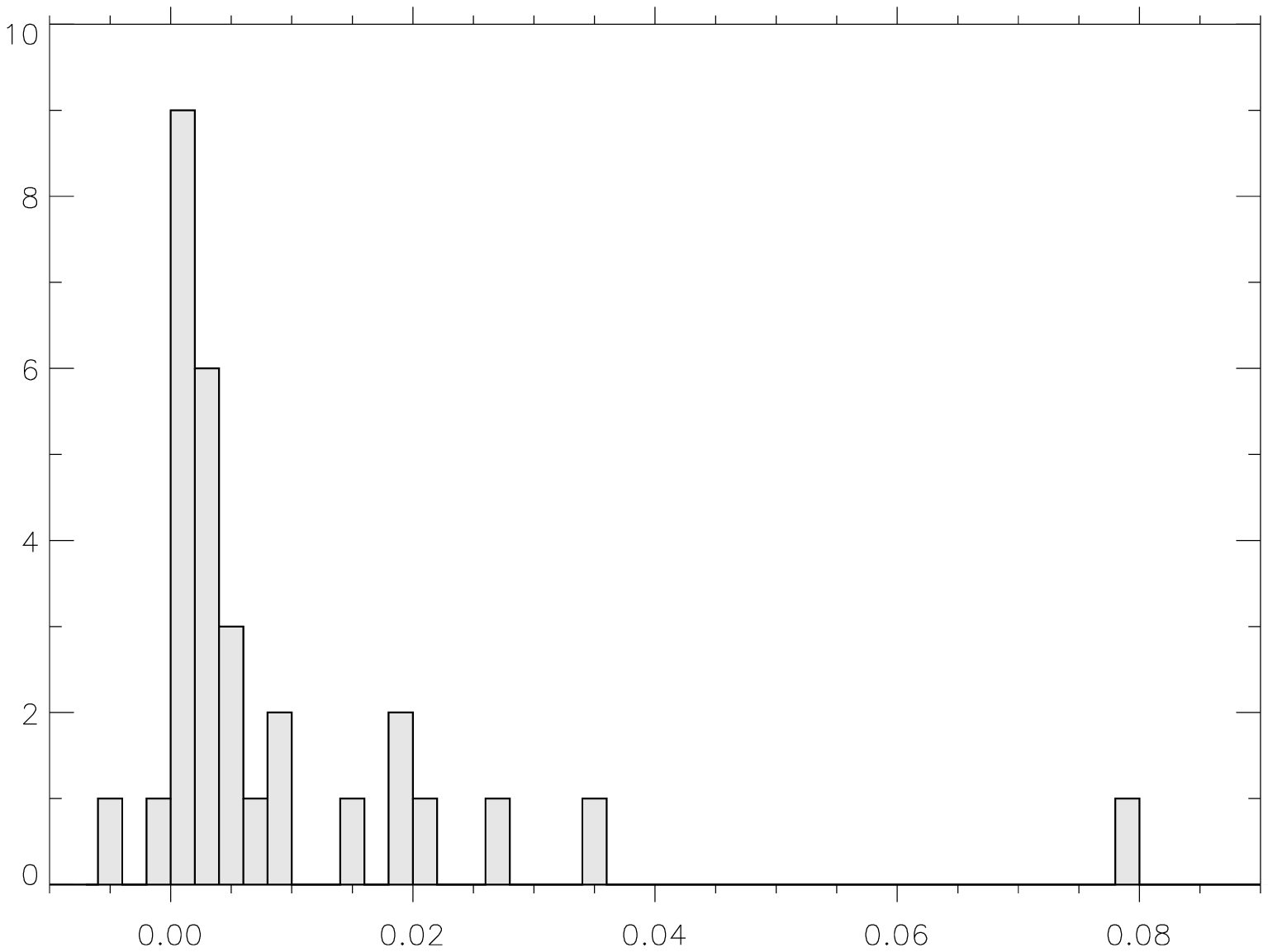,width=12cm,height=8cm,clip=}}

\noindent
Figure 1.~Histogram of values of $\delta\eta_{\rm eddy}$,
amounts of change in magnetic eddy diffusivity due to introduction
of weak time periodicity into the flow.

\vskip6mm
\noindent
b) How does minimal magnetic eddy diffusivity depend
on how kinetic energy is split between the steady and time-dependent
parts of the flow? Magnetic eddy diffusivity has been evaluated
for three sets of sample flows \rf{realflow} for $\omega=1$.
Profiles of the constituent fields ${\bf U}({\bf x})$, ${\bf V}_c({\bf x})$ and ${\bf V}_s({\bf x})$
are the same in each set, and only amplitudes of these fields are varied
in such a way that \rf{enorm} and \rf{equipart} is satisfied.
(In fact, in order to compute any of the three curves,
a sample flow has been chosen out of those used to construct Fig.~1,
and profiles of its constituent fields have been employed.)

Graphs of $\eta_{\rm eddy}$ versus the ratio
$E_{\rm osc}/E_{\rm total}$ are plotted on Fig.~2.
Though graphs representing different flows are quite different in details,
Fig.~2 reveals a common tendency: a relative increase of energy contained
in the time-dependent part of the flow is in general
accompanied by an overall (although not necessarily monotonous)
increase of magnetic eddy diffusivity.

\pagebreak
\centerline{\psfig{file=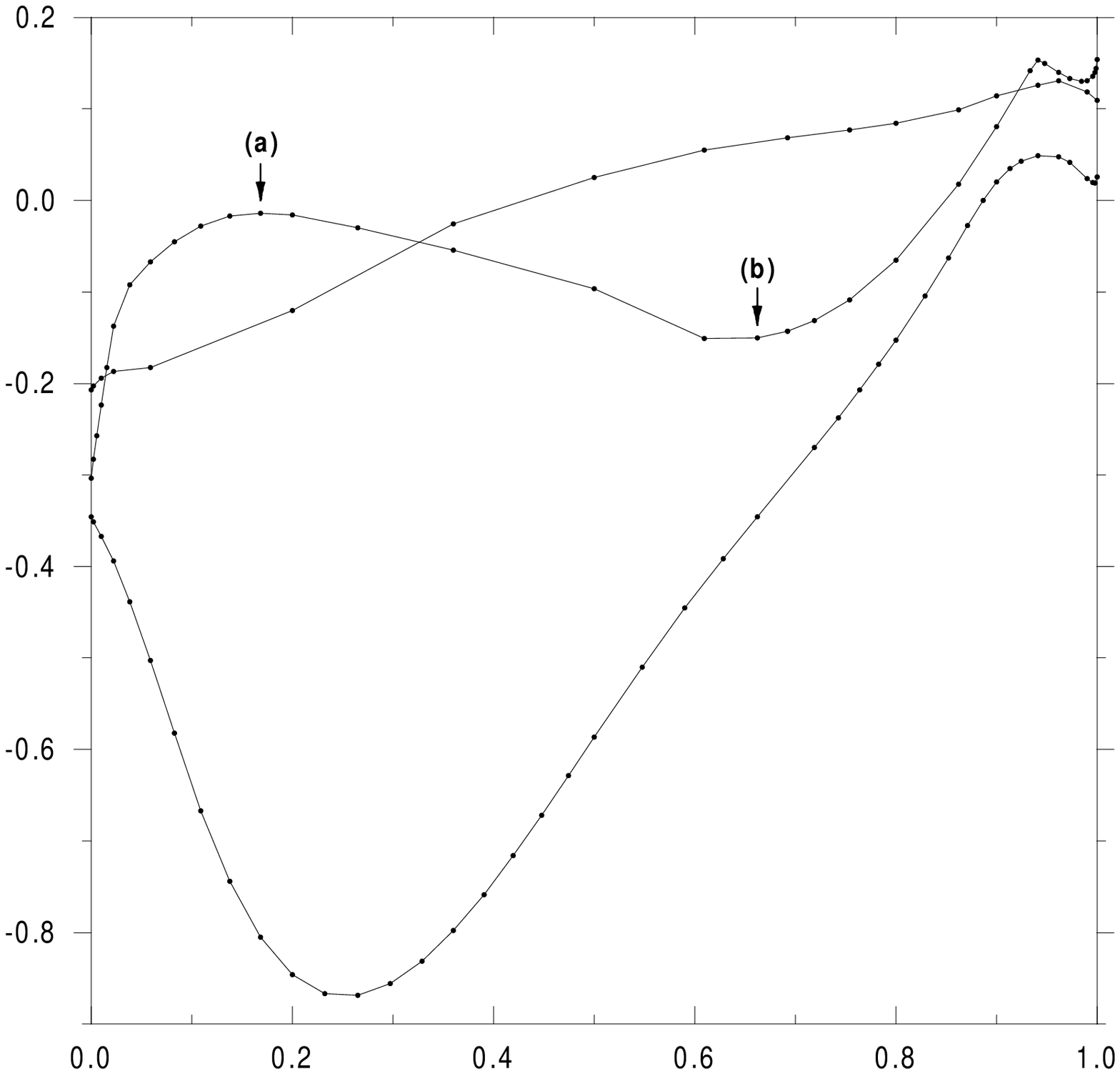,width=135mm,height=11cm,clip=}}

\noindent
Figure 2.~Minimal magnetic eddy diffusivity $\eta_{\rm eddy}$ (vertical axis)
as a function of
the ratio $E_{\rm osc}/E_{\rm total}$ (horizontal axis) for three sets
(represented by three curves) of sample flows \rf{realflow} for $\omega=1$.
Dots show computed values of magnetic eddy diffusivity.
Two dots marked by arrows have been obtained for sample flows,
for which results of computations are shown on Fig.~3.

\vskip6mm
Though time periodicity may be expected to enhance chaotic properties of flows
(which are necessary for fast dynamo action),
our results suggest that time dependence such as
(1) is not favourable for generation of magnetic field.
However, dynamos considered here are slow, and therefore this does not
represent a formal contradiction.

\vskip2mm
\noindent
c) How does magnetic eddy diffusivity depend on temporal frequency
of the flow? Computations have been carried out for two flows (Fig.~3),
which have the same profiles of the constituent fields ${\bf U}({\bf x})$,
${\bf V}_c({\bf x})$ and ${\bf V}_s({\bf x})$.
Two curves on Fig.~3, computed for the two flows, are labelled (a) and (b)
in agreement with labelling of the two respective points on Fig.~2
(indicated on Fig.~2 by arrows). Different patterns of behaviour of magnetic
eddy diffusivity are observed, though the two flows differ
only in amplitudes of their steady and time-dependent parts.
When $\omega\to\infty$, magnetic eddy diffusivity approaches finite
limit values which can be significantly higher than
$\min_\omega\eta_{\rm eddy}(\omega)$, as it happens in the case (b)
(in this case the sample flow has a relatively larger time-dependent part of
the flow). The plots show that the minimal value of $\eta_{\rm eddy}$
is achieved at moderate values of temporal frequency, $\omega=O(1)$.

\pagebreak
\centerline{\psfig{file=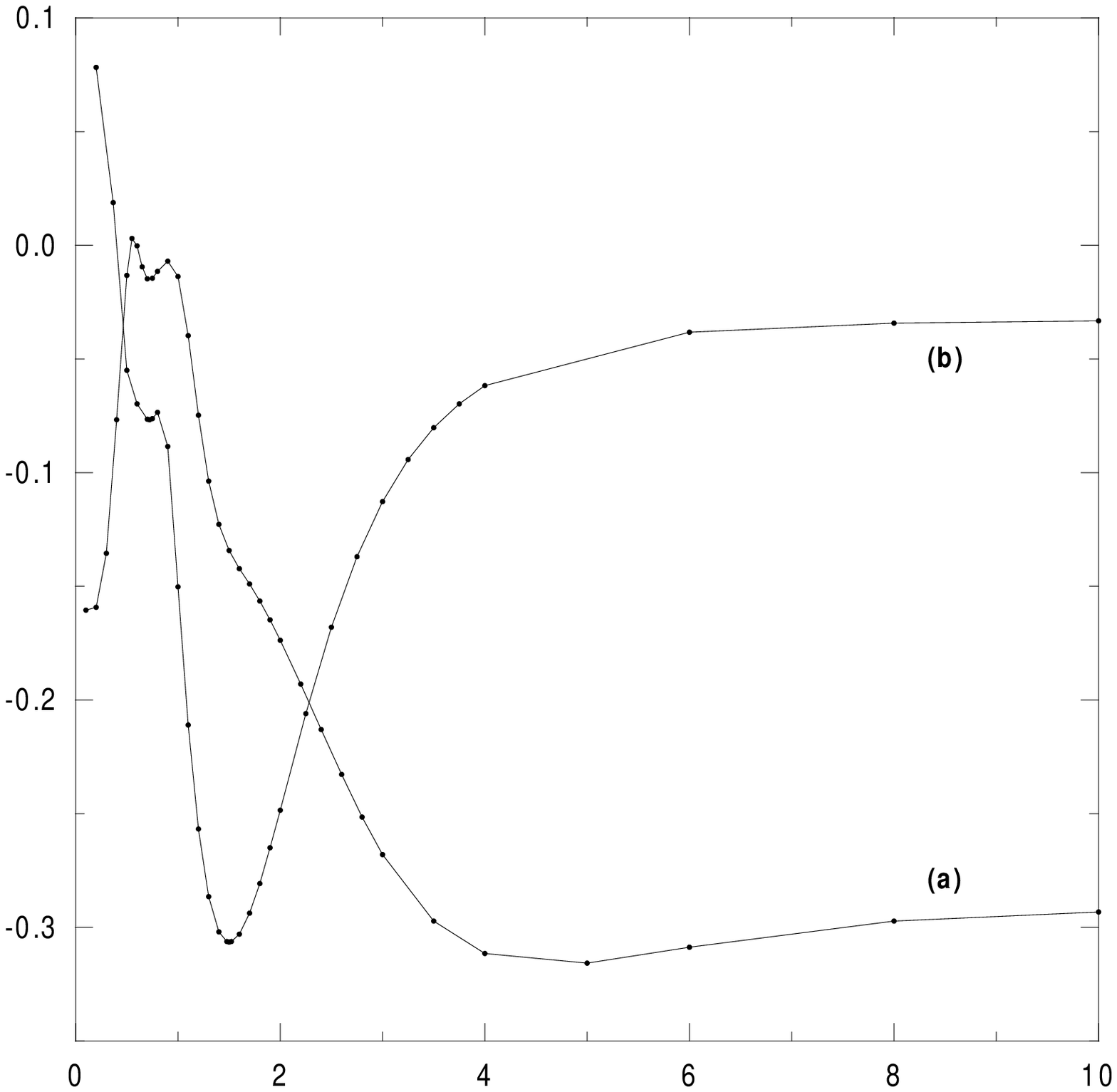,width=135mm,height=11cm,clip=}}

\noindent
Figure 3.~Minimal magnetic eddy diffusivity, $\eta_{\rm eddy}$, (vertical axis)
as a function of temporal frequency of the flow, $\omega$, (horizontal axis).
Dots show computed values of magnetic eddy diffusivity. Curves are labelled
(a) and (b), referring to the labels of the respective points for $\omega=1$ on
Fig.~2.

\vskip6mm
An example of a magnetic mode structure is shown on Fig.~4 and 5.
The leading term in the expansion of
a magnetic mode in the ratio of spatial scales, $\epsilon$, is
$$\tH_0=e^{i\epsilon{\bf q\cdot x}}\left(\ht+\sum_{k=1}^3\ht^k{\bf S}_k({\bf x},t)\right)$$
(see (A6), (A12) and (A22) in the Appendix), where ${\bf S}_k({\bf x},t)$
is the solution to the first auxiliary problem \rf{prob1}, and $\ht$ is
an eigenvector of the problem \rf{oldeigen}-\rf{ortho}.
The fluctuating part of the magnetic mode,
$${\bf G}=\sum_{k=1}^3\ht^k{\bf S}_k({\bf x},t),$$
is shown on Figs.~4 and 5 as surfaces of constant magnetic energy $|{\bf G}|^2$.
The Figures show the magnetic mode generated by that flow, for which the
curve (b) is shown on Fig.~3, for $\omega=1.5$, where the minimum
of magnetic eddy diffusivity is located on this curve.
Cigar-like magnetic structures, seen on the Figures, can be
associated with stagnation points of the flow \rf{realflow}, though
no flux-rope solutions similar to those of Zheligovsky (1993b) and
Galloway and Zheligovsky (1994) are available for time-dependent flows.\break~

\pagebreak
\centerline{\psfig{file=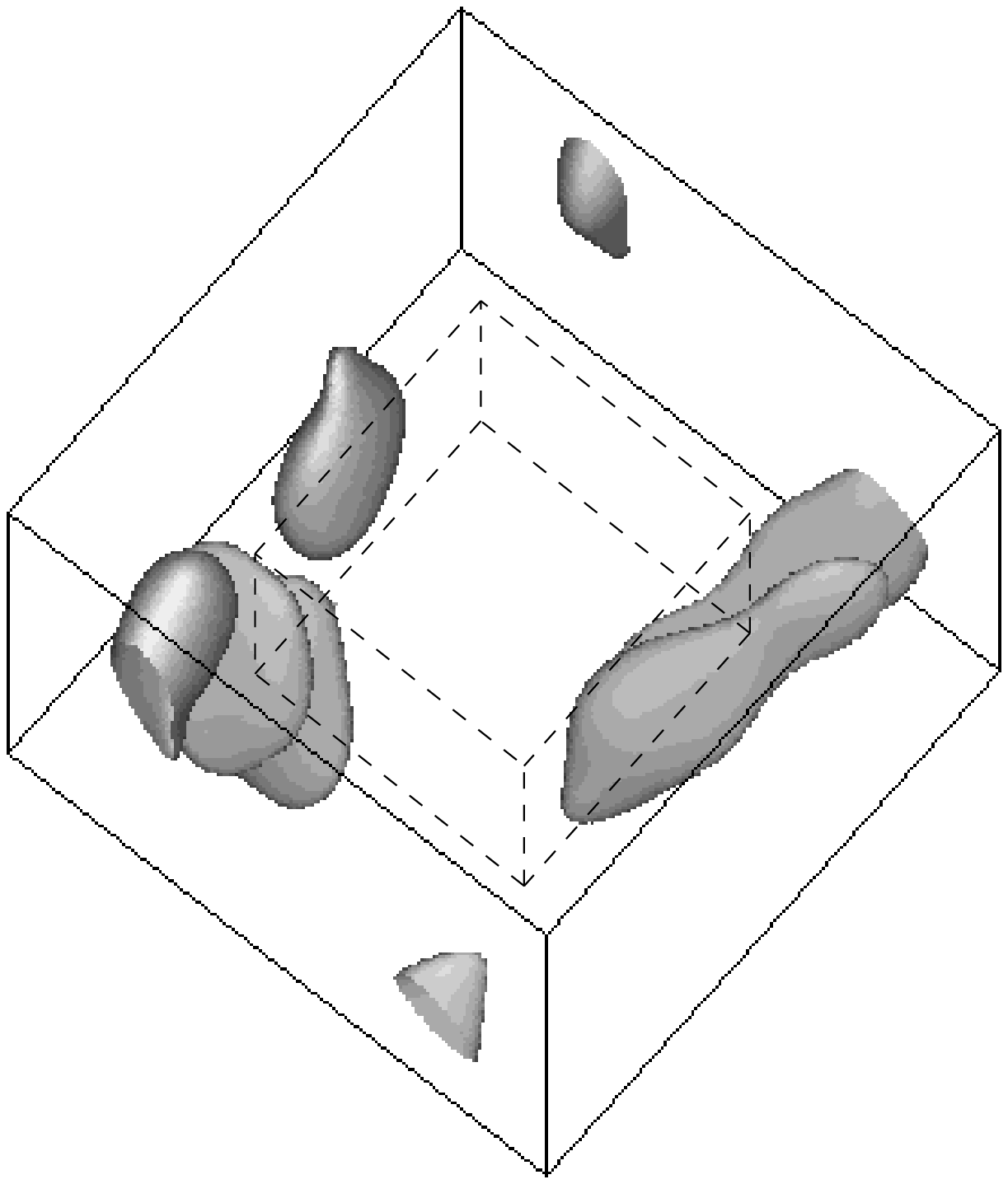,width=7cm,height=67mm,clip=}\hspace{1cm}
\psfig{file=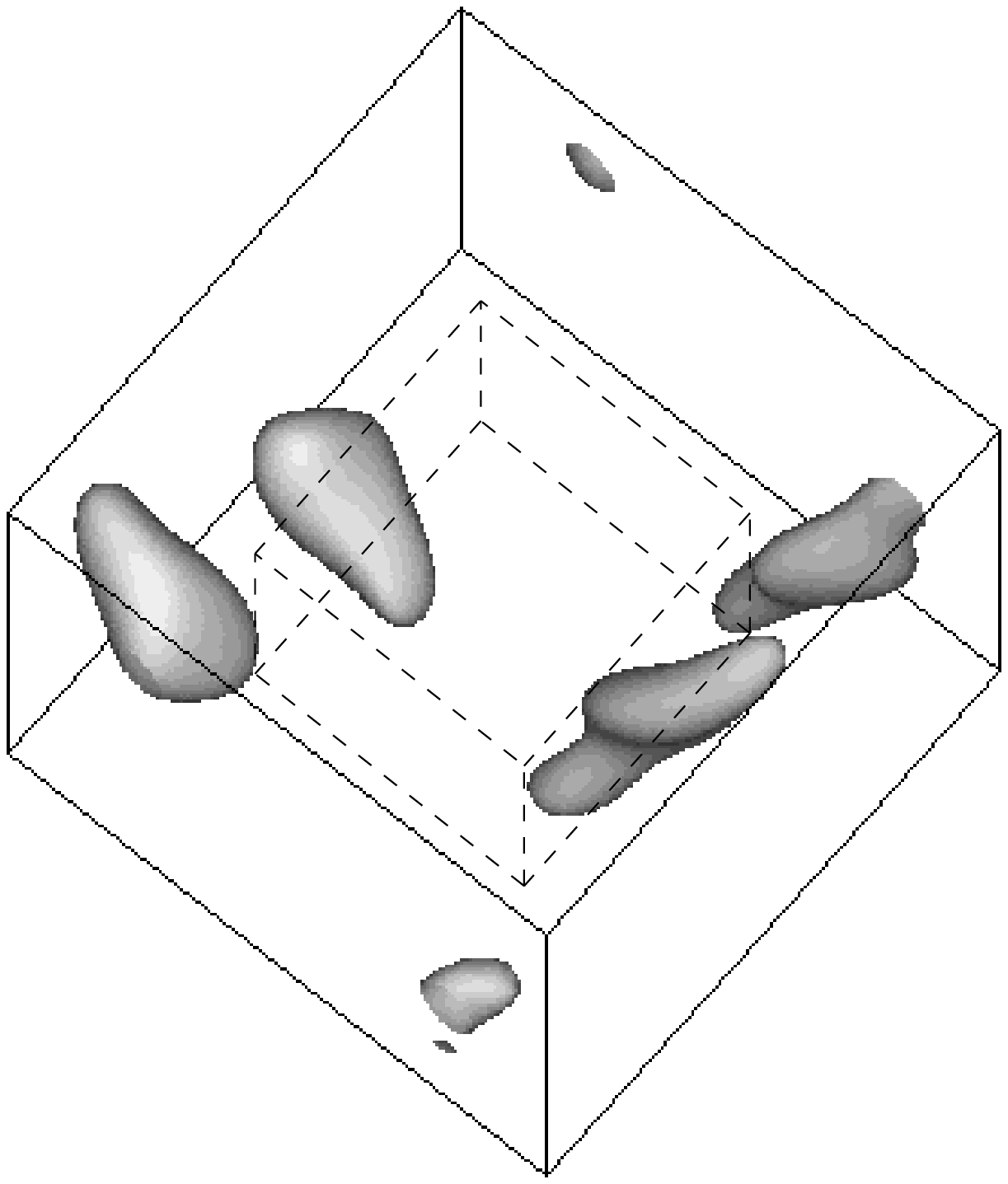,width=7cm,height=67mm,clip=}}

\vspace*{-7mm}
$t=0$\hspace{7cm}$t={T\over2}$

\vspace*{3mm}
\centerline{\psfig{file=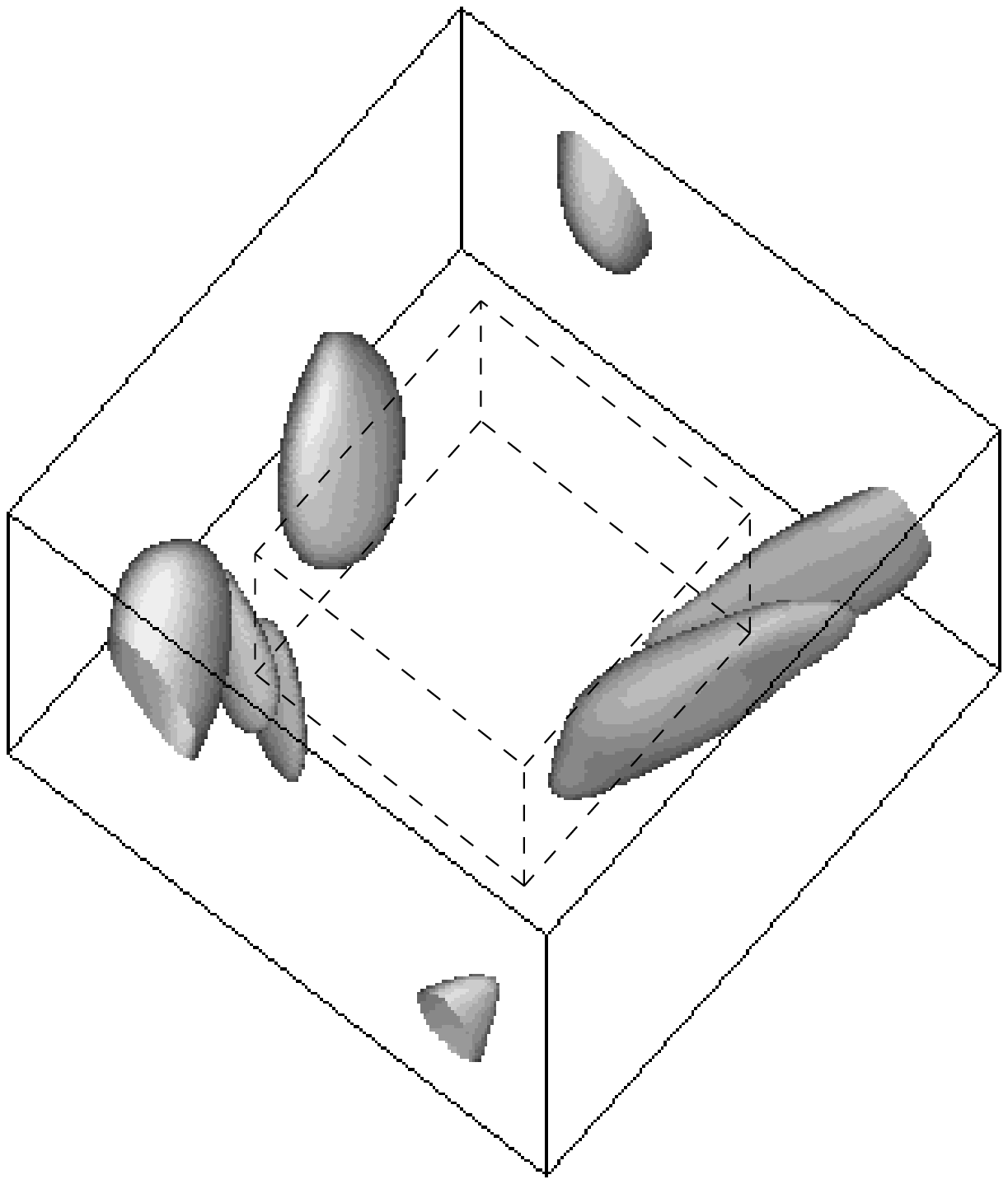,width=7cm,height=67mm,clip=}\hspace{1cm}
\psfig{file=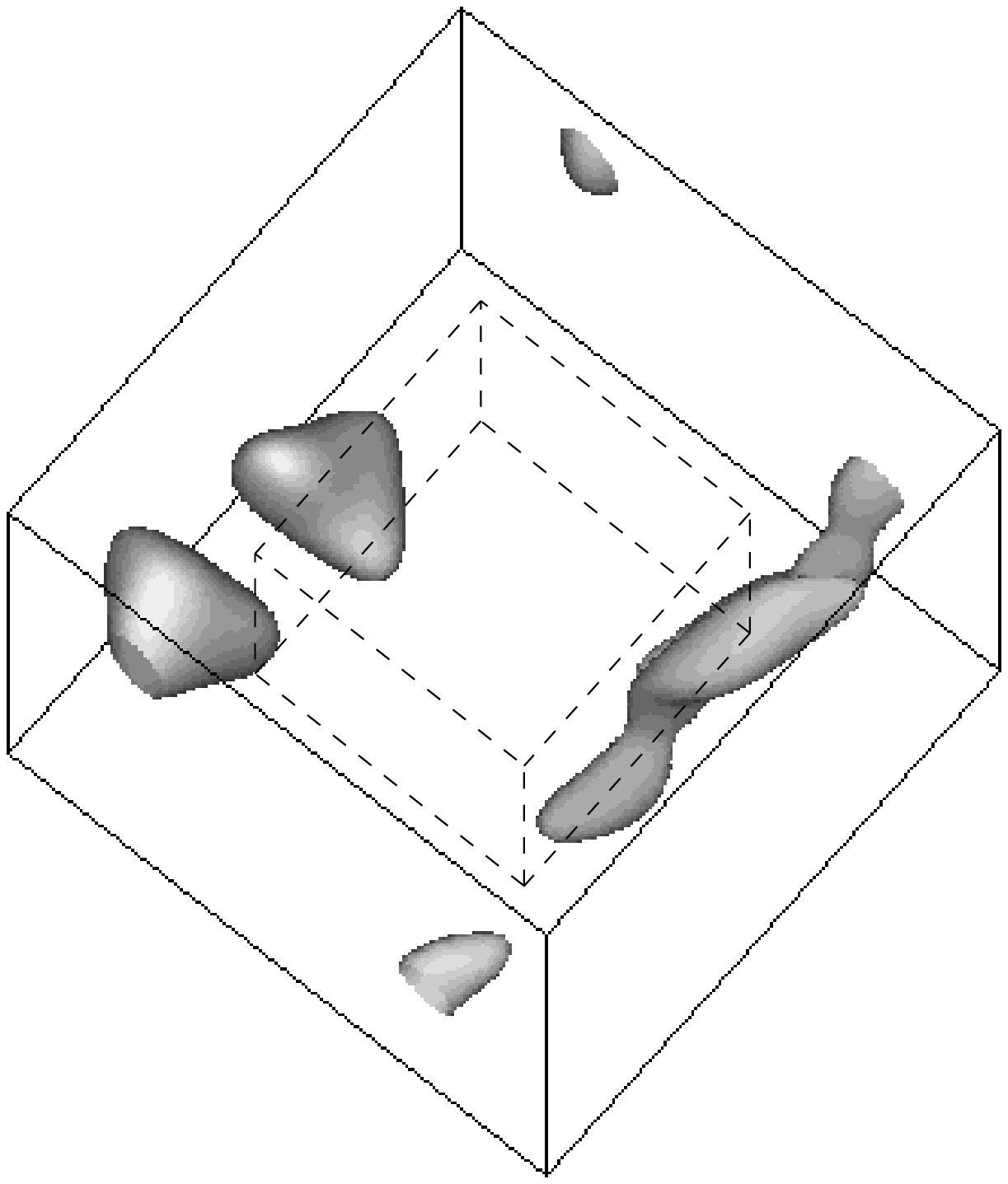,width=7cm,height=67mm,clip=}}

\vspace*{-7mm}
$t={T\over6}$\hspace{7cm}$t={2\over3}T$

\vspace*{3mm}
\centerline{\psfig{file=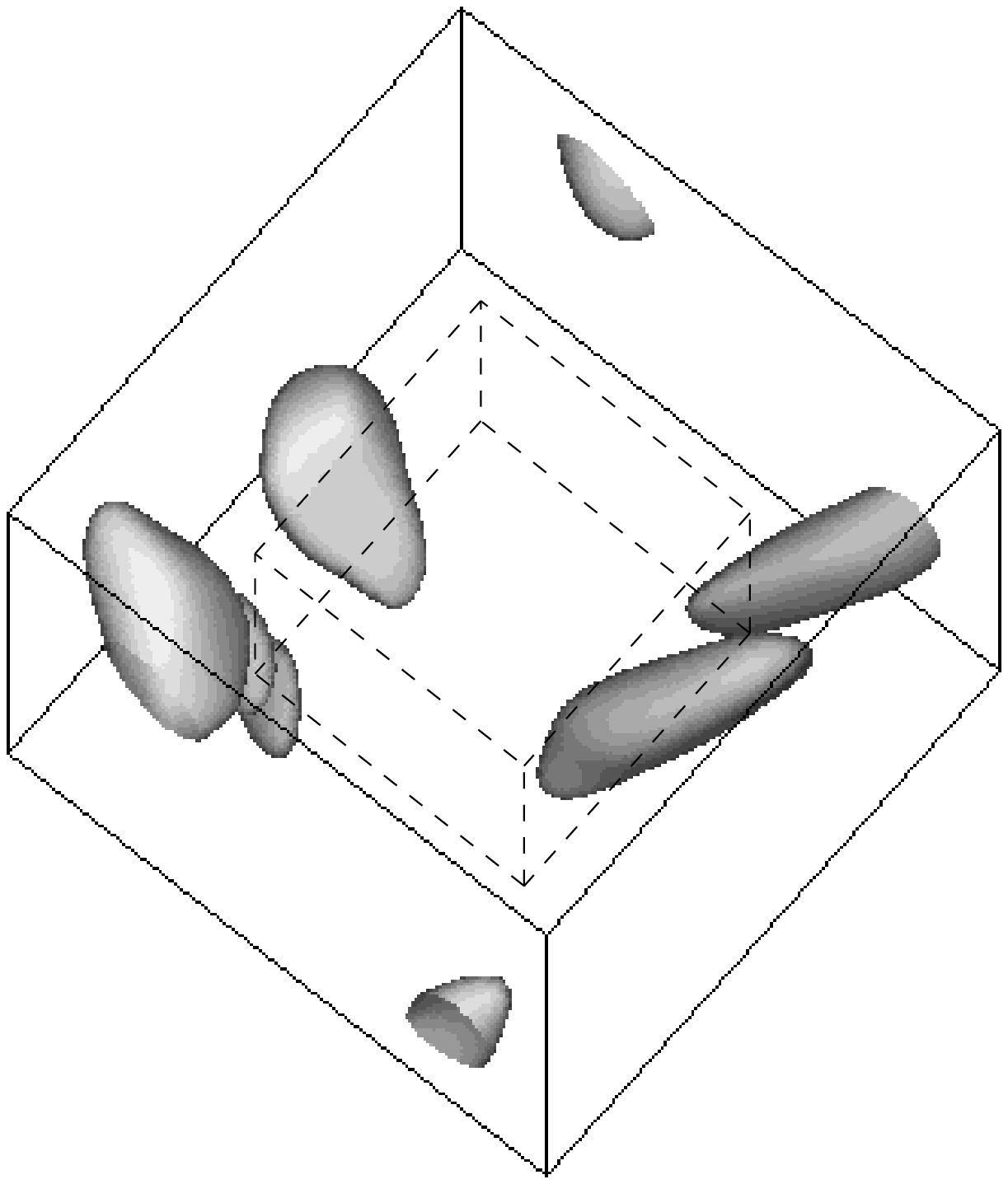,width=7cm,height=67mm,clip=}\hspace{1cm}
\psfig{file=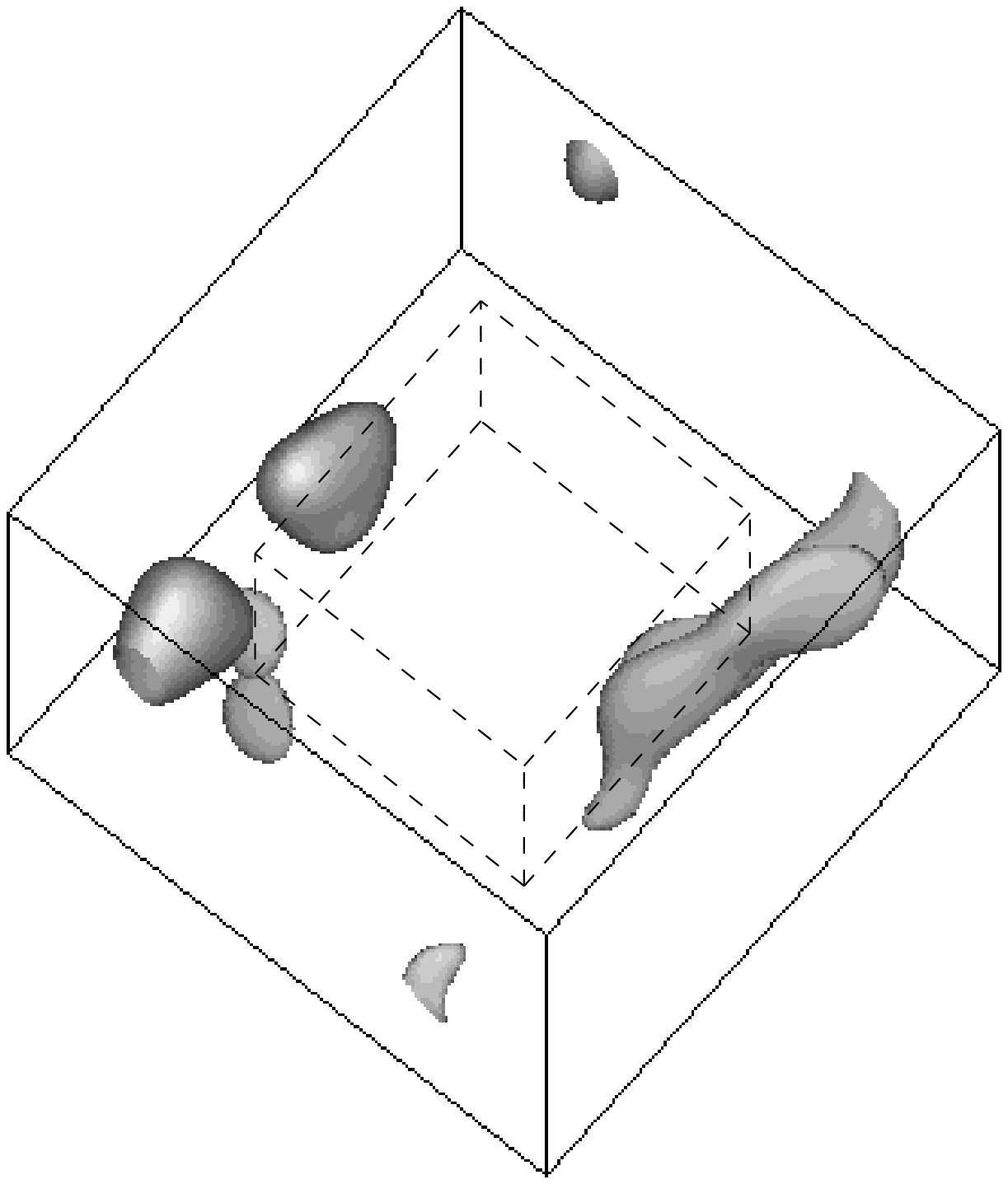,width=7cm,height=67mm,clip=}}

\vspace*{-7mm}
$t={T\over3}$\hspace{7cm}$t={5\over6}T$

\vspace*{5mm}
\noindent
Figure 4.
Isosurfaces of magnetic energy $|{\bf G}|^2$ of the fluctuating part
of a magnetic mode at the level of 40\% of the maximal energy.
Snapshots of a flow periodicity cube (drawn in solid lines) step $T/6$ are
presented. The lower vertex is at the point ($-\pi/2,-\pi/2,-\pi/2$).
Dashed lines show the elementary cube of the flow stagnation points mesh
$(m_1\pi,m_2\pi,m_3\pi)$.

\pagebreak
\centerline{\psfig{file=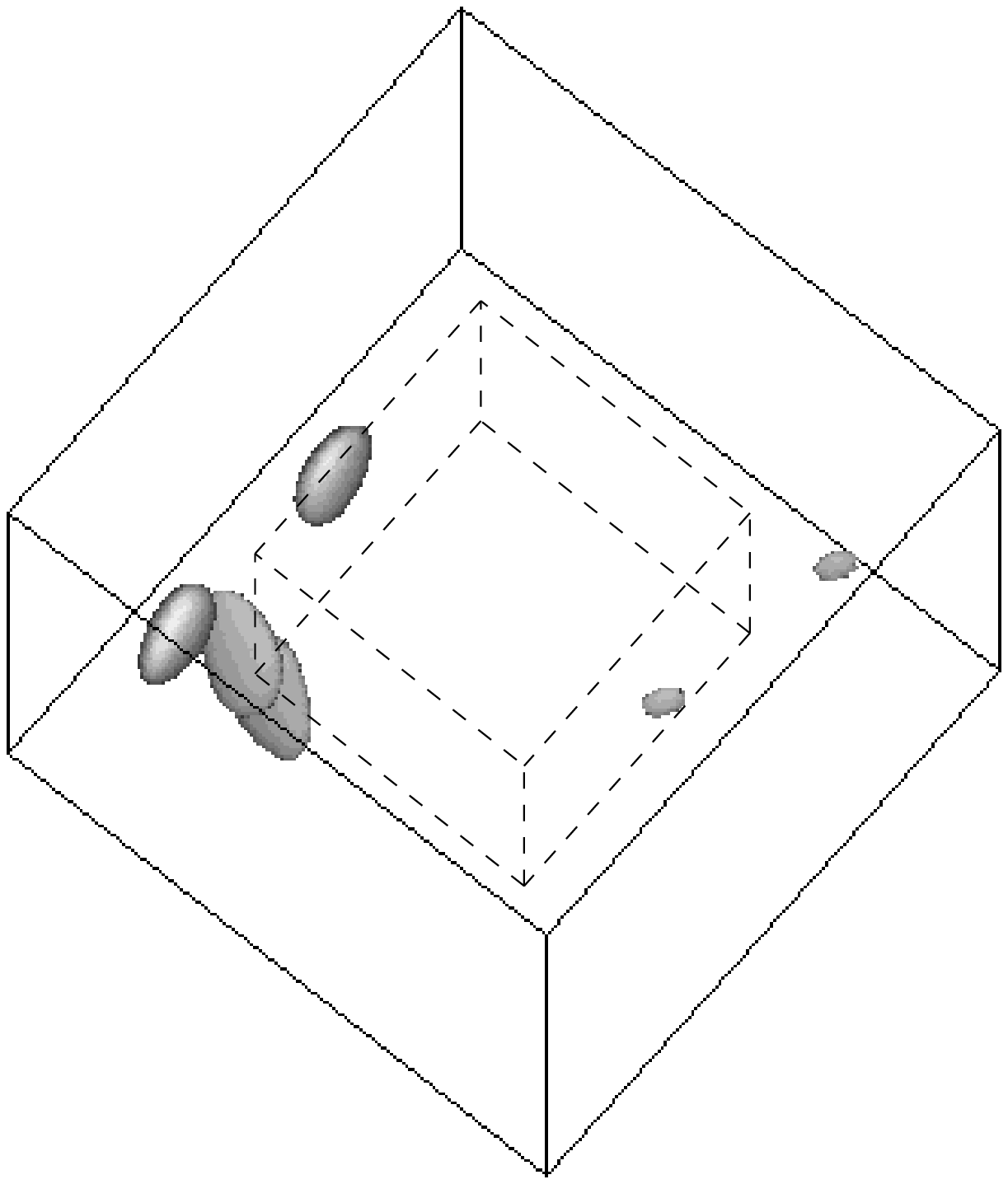,width=7cm,height=67mm,clip=}\hspace{1cm}
\psfig{file=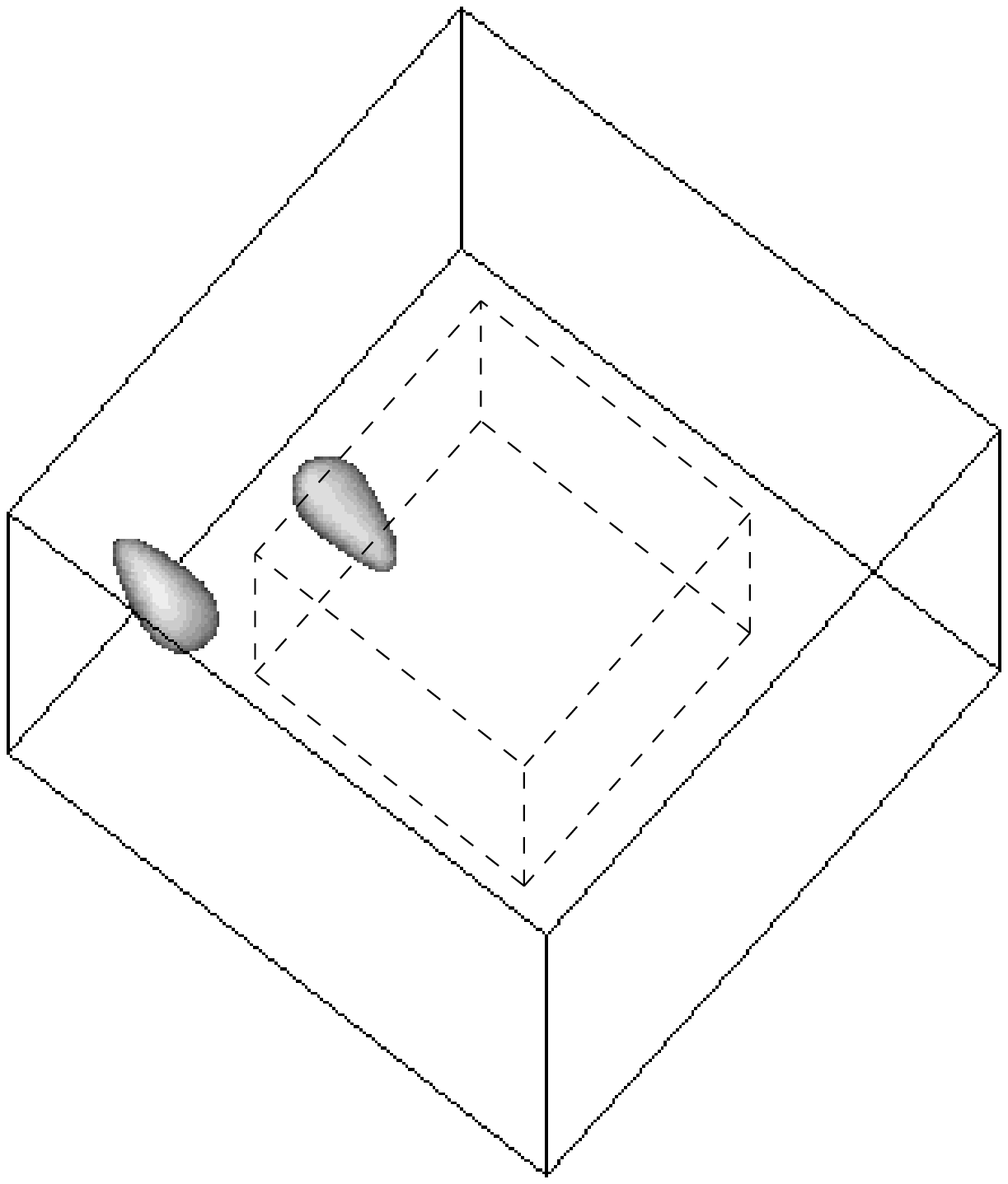,width=7cm,height=67mm,clip=}}

\vspace*{-7mm}
$t=0$\hspace{7cm}$t={T\over2}$

\vspace*{3mm}
\centerline{\psfig{file=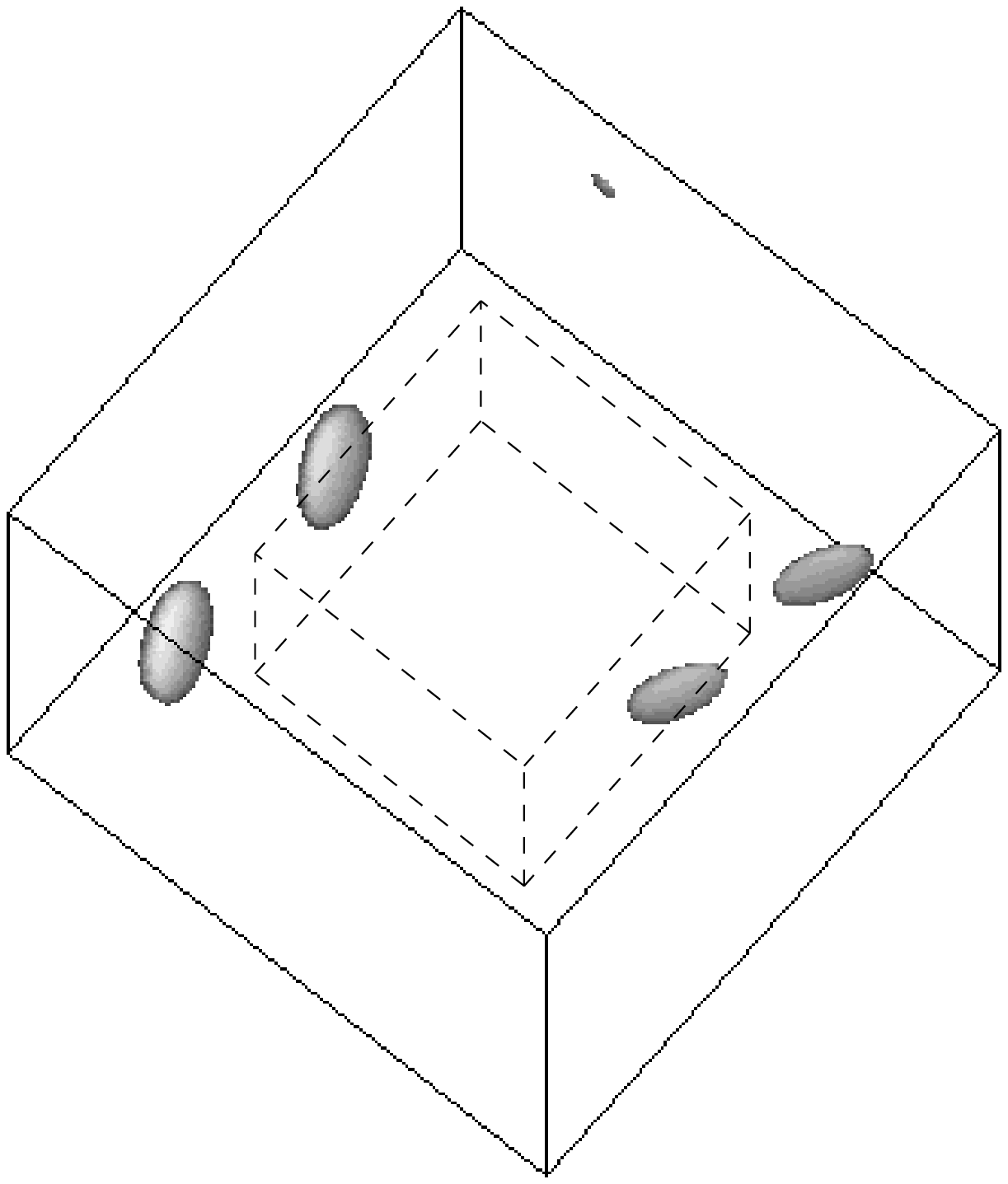,width=7cm,height=67mm,clip=}\hspace{1cm}
\psfig{file=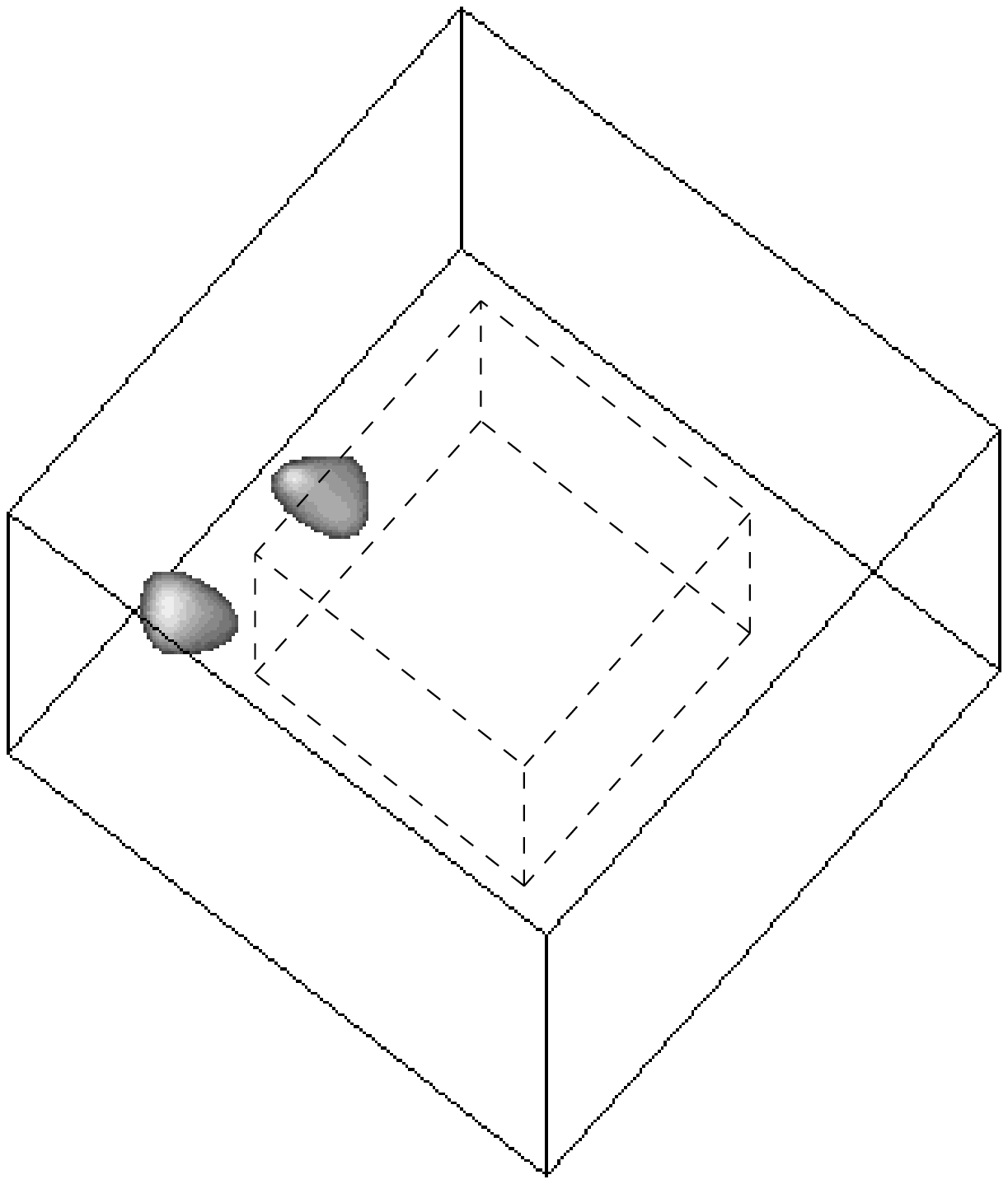,width=7cm,height=67mm,clip=}}

\vspace*{-7mm}
$t={T\over6}$\hspace{7cm}$t={2\over3}T$

\vspace*{3mm}
\centerline{\psfig{file=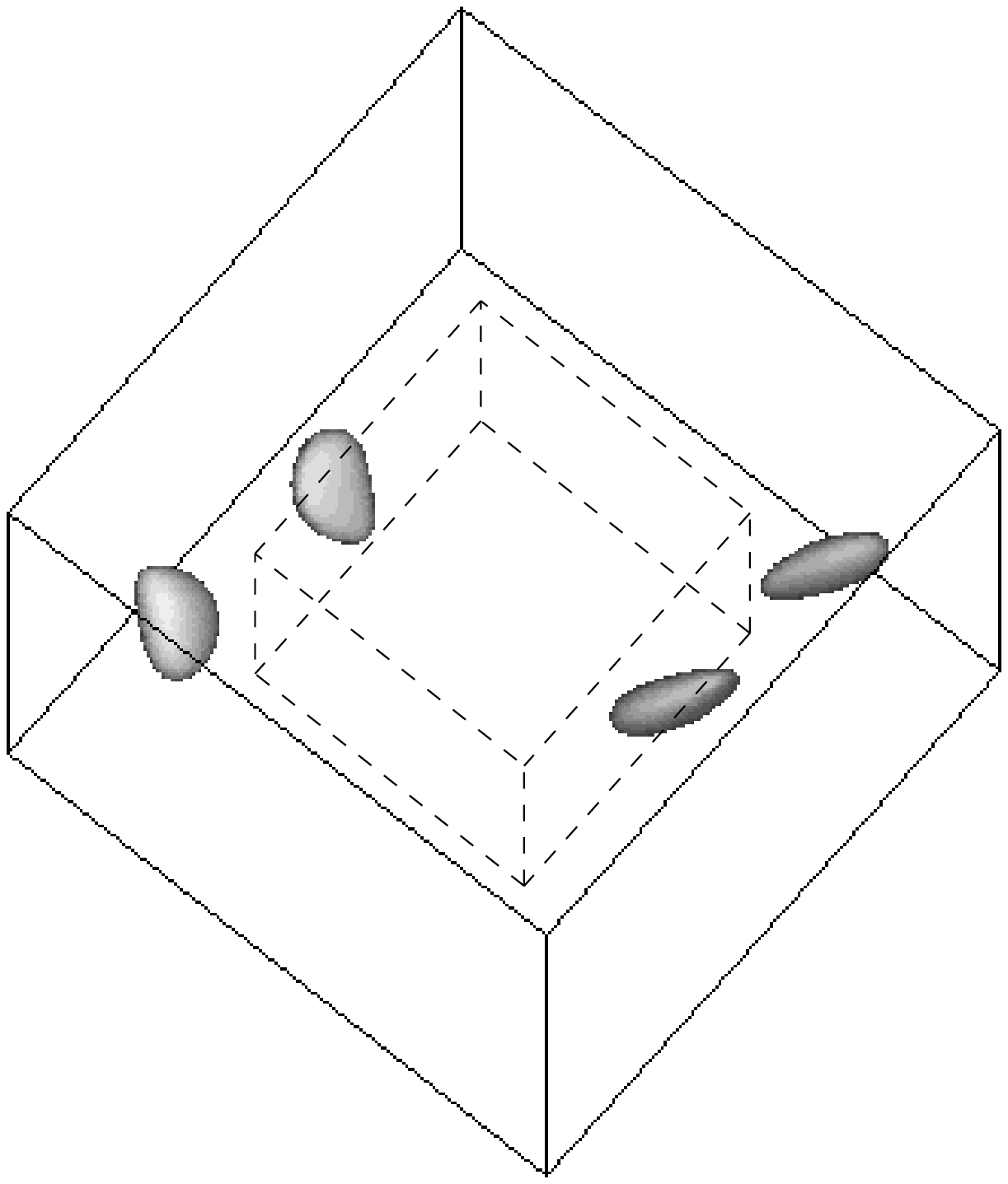,width=7cm,height=67mm,clip=}\hspace{1cm}
\psfig{file=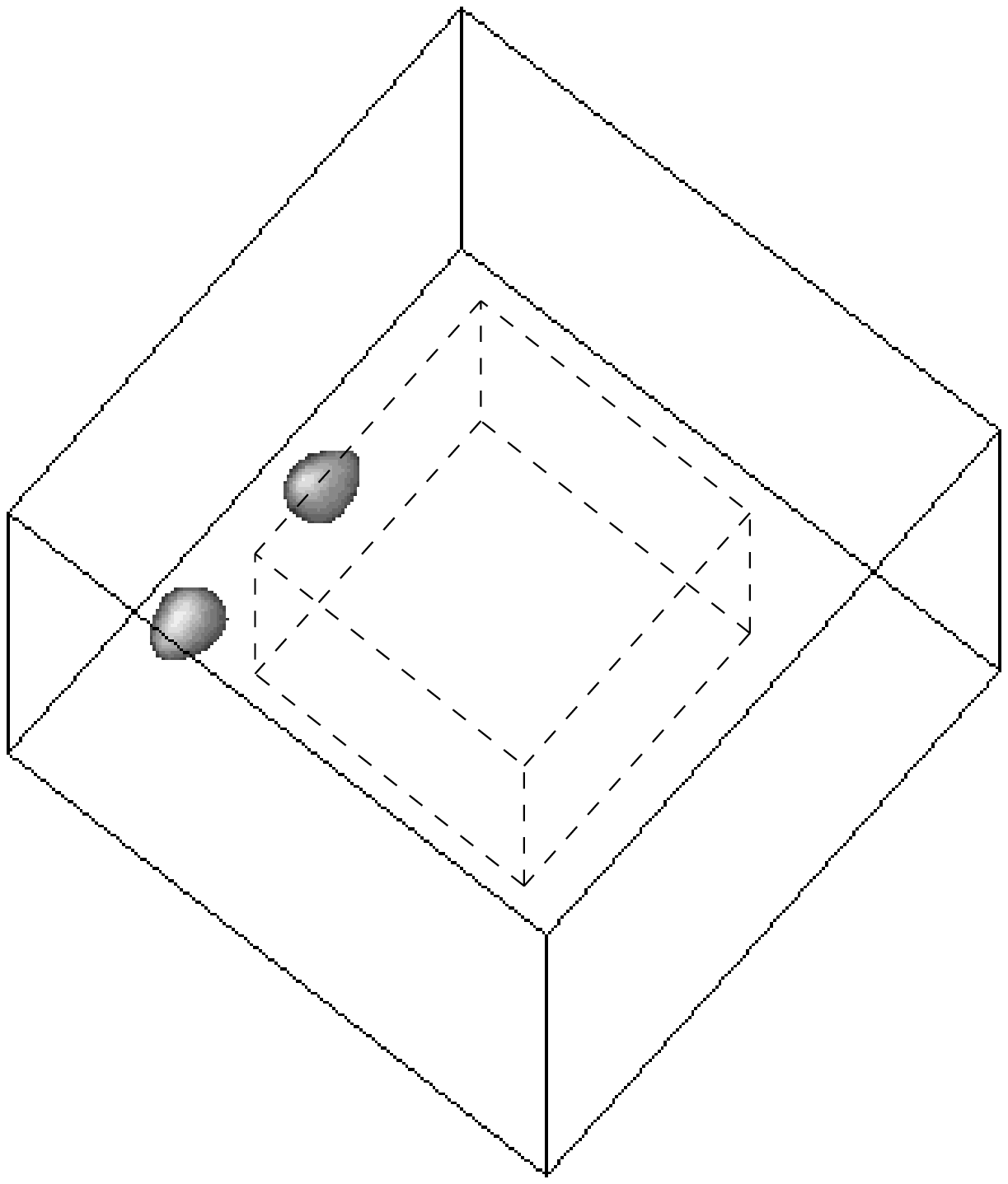,width=7cm,height=67mm,clip=}}

\vspace*{-7mm}
$t={T\over3}$\hspace{7cm}$t={5\over6}T$

\vspace*{8mm}
\noindent
Figure 5.
Same as Fig.~4, but
isosurfaces of magnetic energy of the fluctuating part
of the magnetic mode at the level of 75\% of the maximal energy are shown.

\pagebreak
\noindent
(Note, that due to parity invariance \rf{parity} and space periodicity of
the flow with the period $2\pi$ along each Cartesian axis, the flow velocity
vanishes at the points $(m_1\pi,m_2\pi,m_3\pi)$, where $m_1,m_2,m_3$
are arbitrary integers.) The sharpness and the shape of the cigar structures
vary considerably within the time period.

\bigskip
\noindent
{\bf 3. Magnetic eddy diffusivity of flows \rf{realflow} in the high frequency limit}

\bigskip
Figure 3 suggests, that in the limit $\omega\to\infty$ a contribution to
magnetic eddy diffusivity from the small-scale dynamics of a flow \rf{realflow}
is finite. The limit of high frequencies is studied in this Section.

It is convenient to express the flow \rf{realflow} in the form
\begin{equation}
{\bf v}({\bf x},t)={\bf U}({\bf x})+\sqrt{\omega}
\left({\bf V}({\bf x})e^{i\omega t}+\overline{\bf V}({\bf x})e^{-i\omega t}\right),
\label{imflow}\end{equation}
where ${\bf V}=({\bf V}_c+i{\bf V}_s)/2$.
Denote by $\cal L$ the parabolic magnetic induction operator, acting
in the space of $2\pi/\omega$-periodic small-scale fields ${\bf F}({\bf x},t)$
with a zero mean ($\la{\bf F}\ra=0$):
$${\cal L}{\bf F}\equiv-{\partial{\bf F}\over\partial t}+
\eta\nabla^2{\bf F}+\nabla\times({\bf v}\times{\bf F}).$$
Consider Fourier series of ${\cal L}{\bf F}$:
$${\cal L}{\bf F}=\sum_{j=-\infty}^\infty{\cal L}_j{\bf F}e^{ij\omega t}.$$
For the flow \rf{imflow}
\begin{equation}
{\cal L}_j{\bf F}=-ij\omega{\bf F}_j+\eta\nabla^2{\bf F}_j+\nabla\times({\bf U}\times{\bf F}_j)
+\sqrt{\omega}\left({\bf V}\times{\bf F}_{j-1}+\overline{\bf V}\times{\bf F}_{j+1}\right),
\label{Lcomp}\end{equation}
where ${\bf F}_j$ are temporal Fourier components of $\bf F$:
$${\bf F}({\bf x},t)=\sum_{j=-\infty}^\infty{\bf F}_j({\bf x})e^{ij\omega t}.$$

In terms of the Fourier components, the first auxiliary problem takes the form
\begin{equation}
{\cal L}_j{\bf S}_k=-{\partial\over\partial x_k}\left(\delta^j_0{\bf U}+
\sqrt{\omega}\left(\delta^j_1{\bf V}+\delta^j_{-1}\overline{\bf V}\right)\right),
\label{Seq}\end{equation}
where $\delta$ is the Kronecker symbol, and the second one --
\begin{equation}
{\cal L}_j\Gmm_{m,k}=-2\eta{\partial{\bf S}_{j,k}\over\partial x_m}
+{\bf U}^m\left({\bf S}_{j,k}+\delta^j_0{\bf e}_k\right)-{\bf U}{\bf S}^m_{j,k}
\label{Geq}\end{equation}
$$+\sqrt{\omega}\left({\bf V}^m\left({\bf S}_{j-1,k}+\delta^j_1{\bf e}_k\right)
+\overline{\bf V}^m\left({\bf S}_{j+1,k}+\delta^j_{-1}{\bf e}_k\right)
-{\bf V}{\bf S}^m_{j-1,k}-\overline{\bf V}{\bf S}^m_{j+1,k}\right),$$
where ${\bf S}_{j,k}$ are Fourier components of ${\bf S}_k$:
$${\bf S}_k({\bf x},t)=\sum_{j=-\infty}^\infty{\bf S}_{j,k}({\bf x})e^{ij\omega t}.$$

Assuming a power-law asymptotic behaviour of the Fourier components of ${\bf S}_k$
and $\Gmm_{m,k}$, one finds
$${\bf S}_{j,k}=\omega^{-|j|/2}{\bf s}_{j,k}+O(\omega^{-(|j|+1)/2})$$
to be consistent with \rf{Lcomp} and \rf{Seq}, and
$$\Gmm_{j,m,k}=\omega^{-|j|/2}\gmm_{j,m,k}+O(\omega^{-(|j|+1)/2})$$
is consistent with \rf{Lcomp} and \rf{Geq}.
Here $\Gmm_{j,m,k}$ are Fourier components of $\Gmm_{m,k}$:
$$\Gmm_{m,k}({\bf x},t)=\sum_{j=-\infty}^\infty\Gmm_{j,m,k}({\bf x})e^{ij\omega t}.$$
Leading terms of the Fourier series of ${\bf S}_k$ and $\Gmm_{m,k}$ satisfy
$$\eta\nabla^2{\bf s}_{0,k}+
\nabla_{\bf x}\times\left(2{\rm Re}(\overline{\bf V}\times{\bf s}_{1,k})
+{\bf U}\times{\bf s}_{0,k}\right)=-{\partial{\bf U}\over\partial x_k},$$
\vspace*{-8mm}
\begin{equation}
\label{Slim}\end{equation}
\vspace*{-8mm}
$${\bf s}_{1,k}=-i\left({\partial{\bf V}\over\partial x_k}
+\nabla_{\bf x}\times({\bf V}\times{\bf s}_{0,k})\right);$$
$$\eta\nabla^2\gmm_{0,m,k}+
\nabla_{\bf x}\times\left(2{\rm Re}(\overline{\bf V}\times\gmm_{1,m,k})
+{\bf U}\times\gmm_{0,m,k}\right)$$
\begin{equation}
=-2\eta{\partial{\bf s}_{0,k}\over\partial x_m}
+2{\rm Re}\left(\overline{\bf V}^m{\bf s}_{1,k}-\overline{\bf V}{\bf s}^m_{1,k}\right),
\label{Glim}\end{equation}
$$\gamma_{1,m,k}=-i\left({\bf V}^m({\bf s}_{0,k}+{\bf e}_k)-{\bf V}{\bf s}^m_{0,k}
+\nabla_{\bf x}\times({\bf V}\times\gamma_{0,m,k})\right).$$
Hence, for $\omega\to\infty$ coefficients of the eddy diffusivity tensor are
$${\bf D}_{m,k}=2{\rm Re}\la\overline{\bf V}\times\gamma_{1,m,k}\ra
+\la{\bf U}\times\gamma_{0,m,k}\ra+O(\omega^{-1/2}).$$

The limit values of ${\bf D}_{m,k}$ are not affected by phase shifts:
\rf{Slim} and \rf{Glim} imply that if $\bf V$ is modified to become
$e^{i\alpha}\bf V$, where $\alpha$ is a real constant, then
${\bf s}_{\pm 1,k}$ changes to $e^{\pm i\alpha}{\bf s}_{\pm 1,k}$,
$\gamma_{\pm 1,m,k}$ -- to $e^{\pm i\alpha}\gamma_{\pm 1,m,k}$,
and ${\bf s}_{0,k}$, $\gamma_{0,m,k}$ and ${\bf D}_{m,k}$ remain unaltered.
${\bf s}_{0,k}$ and $\gamma_{0,m,k}$ are real;
if also $\bf V$ is real (i.e. if ${\bf V}_s=\bf 0$), then
${\bf s}_{\pm 1,k}$ and $\gamma_{\pm 1,m,k}$ are imaginary, and hence
$${\rm Re}(\overline{\bf V}\times{\bf s}_{1,k})={\rm Re}(\overline{\bf V}\times\gamma_{1,m,k})=0.$$
Therefore, if Im${\bf V}=\bf 0$, then in the limit $\omega\to\infty$
a contribution from the time-periodic part of the flow \rf{imflow} vanishes.
Combined together, these two observations imply that a non-zero contribution
from the time-periodic part of \rf{imflow} requires linear independence
of vector fields ${\bf V}_c$ and ${\bf V}_s$.

The same value of molecular viscosity, $\eta=0.1$, has been employed
in simulations. Short-scale magnetic modes can be analyzed
in the limit $\omega\to\infty$ along the same lines, as above: their
temporal Fourier components exhibit the same power-law asymptotical behaviour,
as solutions to the auxiliary problems.
The limit small-scale magnetic modes turn out to be eigenfunctions of the linear
operator defined by the left-hand sides of \rf{Slim} and \rf{Glim}.
It has been verified that for each sample flow \rf{realflow},
for which a limit magnetic eddy diffusivity value is reported here,
the limit magnetic induction operator acting in the space of $2\pi$-periodic small-scale
zero-mean magnetic fields has no eigenvalues with a positive real part,
i.e.~for the employed flows $\eta=0.1$ is above the threshold
for the onset of generation of small-scale magnetic fields
in the limit of high temporal frequencies.

We carried out the following numerical experiments:

\vskip 2mm
\noindent
a) Limit values of magnetic eddy diffusivities have been computed using
\rf{Slim} and \rf{Glim} for those three sets of sample flows \rf{realflow}
(see Fig.~6), for which plots of Fig.~2 have been constructed for $\omega=1$.
The left-most points (for $E_{\rm osc}=0$) of two respective curves on Figs.~2
and 6, obtained for the same set of sample flows, show the same value;
this can be used to identify pairs of the respective curves.
Like for finite frequency (see Section 2),
a relative increase of energy contained in the time-dependent part of the flow
is accompanied by an overall increase of magnetic eddy diffusivity,
though the dependence is not necessarily monotonous. However, the influence
of time-dependent parts of flows has now weakened.

\vskip6mm
\centerline{\psfig{file=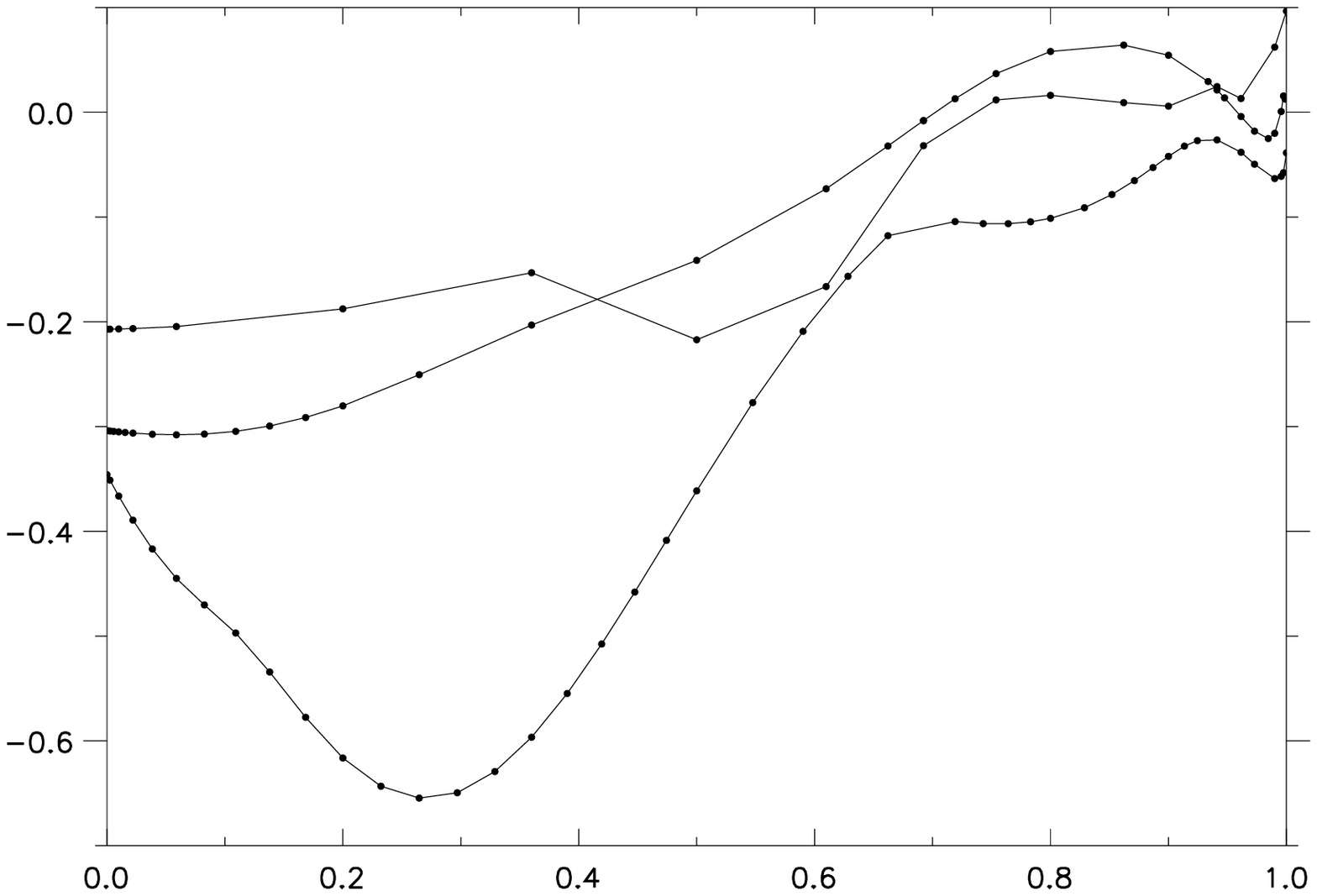,width=135mm,height=11cm,clip=}}

\noindent
Figure 6.~Minimal magnetic eddy diffusivity $\eta_{\rm eddy}$ (vertical axis)
in the limit $\omega\to\infty$ as a function of
the ratio $E_{\rm osc}/E_{\rm total}$ (horizontal axis) for three sets
(represented by three curves) of sample flows \rf{realflow}.
Dots show computed values of magnetic eddy diffusivity.

\pagebreak
\noindent
b) Distribution of minimal magnetic eddy diffusivity for \rf{realflow} has been
examined under the conditions, apparently least favourable for generation --
in the absence of the steady component (i.e. for ${\bf U}=\bf 0$)
and in the limit $\omega\to\infty$. A histogram of limit values of
$\eta_{\rm eddy}$ computed for 45 sample flows,
satisfying \rf{enorm}, \rf{equipart} and ${\bf U}={\bf 0}$, is shown on Fig.~7.
Only in 2 cases out of 45 the limit magnetic eddy diffusivity is negative.

\vskip6mm
\centerline{\psfig{file=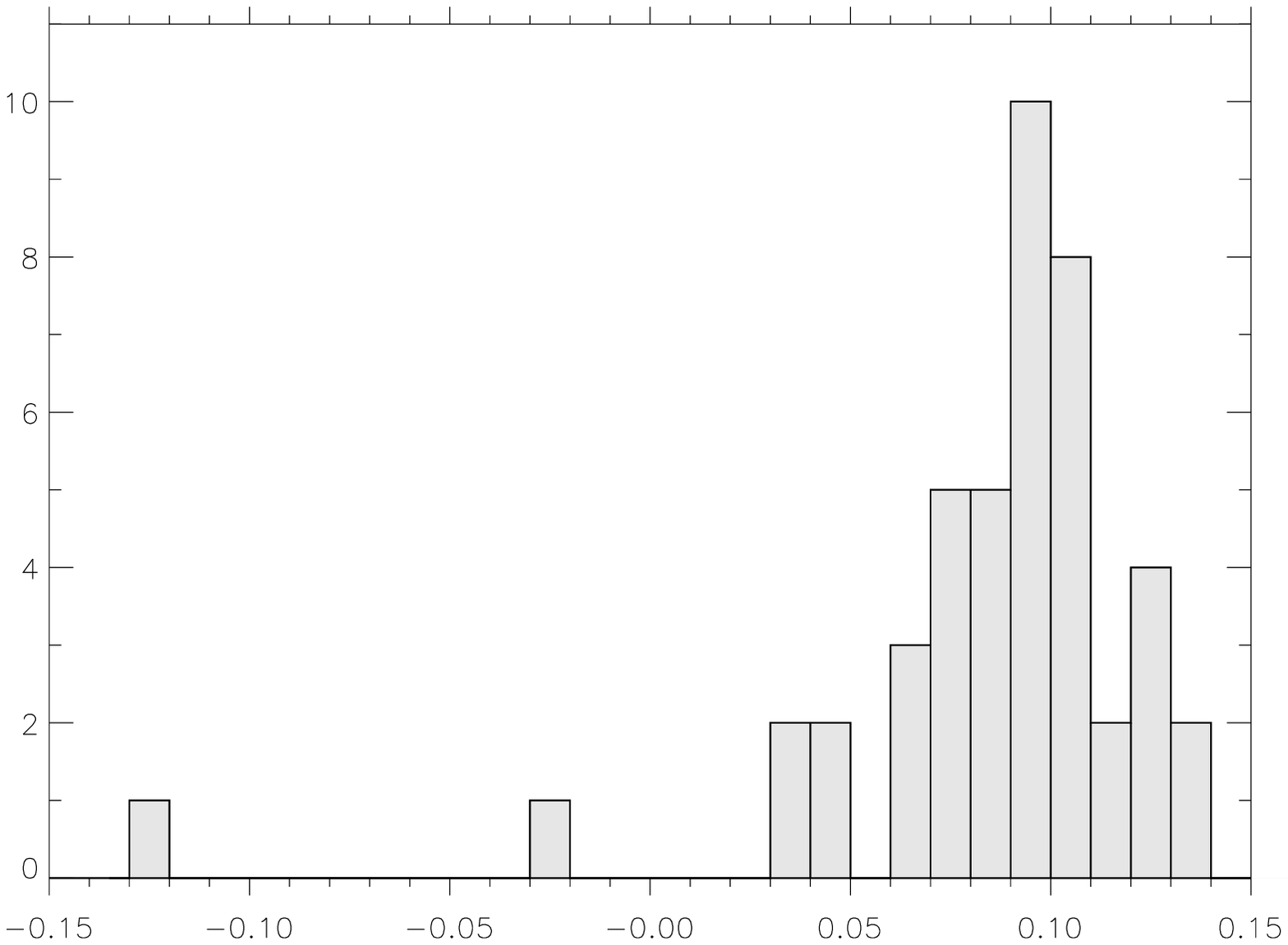,width=12cm,height=8cm,clip=}}

\noindent
Figure 7.~Histogram of minimal magnetic eddy diffusivity values in the limit
\hbox{$\omega\to\infty$} for 45 sample flows satisfying \rf{enorm},
\rf{equipart} and ${\bf U}=\bf 0$.

\vskip6mm
\noindent
{\bf Conclusion}

\bigskip
Generation of magnetic field involving large spatial scales by time-
and space-periodic small-scale parity-invariant flows has been studied.
The aniso\-tropic magnetic eddy diffusivity tensor has been calculated.
A complete expansion of magnetic modes and their growth rates in power series
in the scale ratio has been constructed for flows \rf{realflow}.
Simulations have been carried out for flows \rf{realflow} with random harmonic
composition and an exponentially decaying energy spectrum.
Flows giving rise to negative magnetic eddy diffusivity have been found,
for molecular diffusivity above the instability threshold
for small-scale magnetic field generation. Numerical results suggest
that generation of large-scale magnetic field by time-periodic flows
is less efficient than generation by steady flows of the same energy.
High temporal frequencies may also be unfavourable for generation.
Nevertheless, we have found numerically instances of flows \rf{realflow}
with a vanishing steady component, for which magnetic
eddy diffusivity is negative for $\omega\to\infty$.

\pagebreak
\noindent
{\bf Acknowledgments}

\bigskip
We are grateful to U.~Frisch for extensive discussions. In the course
of preparation of the final version of the paper we have benefited from
the referees' remarks. A part of numerical results were obtained using
computational facilities provided by the program ``Simulations
Interactives et Visualisation en Astronomie et M\'ecanique (SIVAM)''
at Observatoire de la C\^ote d'Azur and by the D\'epartement Sciences
Physiques pour l'Ing\'enieur of CNRS, France. Our research at
Observatoire de la C\^ote d'Azur was supported by the French
Ministry of Education. The investigation was concluded during our stay
at the School of Mathematical Sciences, University of Exeter, UK,
in May -- July 2002.
We are grateful to the Royal Society for their support of the visit.

\bigskip
\noindent
{\bf References}

\mi
Biferale L., Crisanti A., Vergassola M. and Vulpiani A. ``Eddy diffusivity
in scalar transport", {\it Phys. Fluids} {\it 7}, 2725--2734 (1995).

\mi
Brummell N.H., Cattaneo F. and Tobias S.M.
``Nonlinear dynamo action in a time-dependent ABC flow",
in {\it Stellar dynamos: Nonlinearity and chaotic flows.
Proc.~of the International workshop held at Medina del Campo,
Valladolid, Spain, 28-30 September, 1998}
(Eds. M.~N\'u\~nez and A.~Ferriz-Mas). Astr. Soc. of the Pacific,
Conf. series vol.~{\bf 178}, San Francisco, 23--34 (1999).

\mi
Brummell N.H., Cattaneo F. and Tobias S.M.
``Linear and nonlinear dynamo properties of time-dependent ABC flows", Fluid
Dynamics Res. {\bf 28}, 237--265 (2001).

\mi
Childress S. and Gilbert A.D. {\it Stretch, twist, fold: the fast dynamo},
Springer-Verlag, Berlin (1995).

\mi
Galloway D.J. and O'Brian, N.R.
``Numerical calculations of dynamos for ABC and related flows",
in {\it Solar and Planetary Dynamos}
(Eds. M.R.E. Proctor, P.C. Matthews, A.M. Rucklidge)
Cambridge Univ. Press, 105--113 (1993).

\mi
Galloway D.J. and Proctor, M.R.E.
``Numerical calculations of fast dynamos for smooth velocity fields with
realistic diffusion", {\it Nature} {\bf 356}, 691--693 (1992).

\mi
Galloway D.J. and Zheligovsky V.A. ``On a class of non-axisymmetric
flux rope solutions to the electromagnetic induction equation",
{\it Geophys. Astrophys. Fluid Dynamics}, {\bf 76}, 253--264 (1994).

\mi
Lanotte A., Noullez A., Vergassola M. and Wirth A.
``Large-scale dynamo by negative magnetic eddy diffusivities",
{\it Geophys. Astrophys. Fluid Dynamics} {\bf 91}, 131--146 (1999).

\mi
Otani N.F. ``A fast kinematic dynamo in two-dimensional time-dependent flows",
{\it J.~Fluid Mech.} {\bf 253}, 327--340 (1993).

\mi
Ponty Y., Pouquet A., Rom-Kedar A. and Sulem P.L. ``Dynamo action
in a nearly integrable chaotic flow", in {\it Solar and Planetary Dynamos}
(Eds. M.R.E. Proctor, P.C. Matthews, A.M. Rucklidge)
Cambridge Univ. Press, 241--248 (1993).

\mi
Ponty Y., Pouquet A. and Sulem P.L. ``Dynamos in weakly chaotic
two-dimensional flows",
{\it Geophys. Astrophys. Fluid Dynamics} {\bf 79}, 239--257 (1995).

\mi
Zheligovsky V.A. ``Numerical solution of the kinematic dynamo problem
for Beltrami flows in a sphere", {\it J. of Scientific Computing},
{\bf 8}, 41--68 (1993a).

\mi
Zheligovsky V.A. ``A kinematic magnetic dynamo sustained by a Beltrami
flow in a sphere", {\it Geophys. Astrophys. Fluid Dynamics}, {\bf 73},
217--254 (1993b).

\mi
Zheligovsky V.A., Podvigina O.M. and Frisch U.
``Dynamo effect in parity-invariant flow with large and moderate
separation of scales", {\it Geophys. Astrophys. Fluid Dynamics}
{\bf 95}, 227-268 [http://xxx.lanl.gov/abs/nlin.CD/0012005] (2001).

\pagebreak
\noindent
{\bf Appendix. Formal asymptotic decomposition of magnetic modes\\
and of their growth rates for time-periodic flows \rf{realflow}}

\bigskip
We derive complete asymptotic expansions of a magnetic mode and its growth rate
for the flow \rf{realflow}, which it is convenient to express here as
$${\bf v}({\bf x},t)={\bf U}({\bf x})+
{\bf W}({\bf x})e^{i\omega t}+\overline{\bf W}({\bf x})e^{-i\omega t},\eqn{A1}$$
where the notation
$${\bf W}=\sqrt{\omega}{\bf V}\eqn{A2}$$
is used. Our derivation follows that of Zheligovsky\al(2001) for steady flows.

As in the main body of the paper, ${\bf x}\in R^3$ denotes the fast, and
${\bf y}=\epsilon\bf x$ -- the slow spatial variable,
$\la\cdot\ra$ and $\lb\cdot\rb$
denote the mean and the fluctuating part of a vector field, respectively:
$$\la{\bf f({\bf x},{\bf y},t)}\ra\equiv(2\pi)^{-3}\int_{[0,2\pi]^3}{\bf f}({\bf x},{\bf y},t)d{\bf x},\quad
\lb{\bf f}({\bf x},{\bf y},t)\rb\equiv{\bf f}-\la{\bf f}\ra.$$

The following assumptions are made concerning the flow $\bf v$:
$\bf U$ and $\bf W$ are $2\pi$-periodic in
fast Cartesian variables, independent of time and of the slow variables, solenoidal:
$$\nabla\cdot{\bf U}=\nabla\cdot{\bf W}=0,\eqn{A3}$$
and parity-invariant:
$${\bf U}({\bf x})=-{\bf U}(-{\bf x}),\qquad{\bf W}({\bf x})=-{\bf W}(-{\bf x}).\eqn{A4}$$

A magnetic mode ${\bf H}({\bf x},{\bf y},t)$ is a solenoidal \rf{Hsolen}
solution to the Floquet problem \rf{floquet},
which is $2\pi$-periodic in each spatial variable
and has the same temporal period $T=2\pi/\omega$, as the flow.
For the flow (A1), Fourier components of a magnetic mode
$${\bf H}({\bf x},{\bf y},t)=\sum_{j=-\infty}^\infty
{\bf H}_j({\bf x},{\bf y})e^{ij\omega t}$$
satisfy
$$\lambda{\bf H}_j=-ij\omega{\bf H}_j+\eta\nabla^2{\bf H}_j+\nabla\times\left({\bf W}\times{\bf H}_{j-1}
+{\bf U}\times{\bf H}_j+\overline{\bf W}\times{\bf H}_{j+1}\right)\eqn{A5}$$
for all $j$.

A solution to this system of equations is sought in the form of power series
$${\bf H}_j({\bf x},{\bf y})=\sum_{n=0}^\infty({\bf H}_{j,n}({\bf y})+{\bf G}_{j,n}({\bf x},{\bf y}))
\epsilon^n\eqn{A6}$$
and \rf{grate}. In (A6) ${\bf H}_{j,n}$ and ${\bf G}_{j,n}$ are the mean and
the fluctuating part of the respective term of the series: $\la{\bf G}_{j,n}\ra=0$.
Obviously, any term $\tH_n$ of the expansion \rf{mmode} can be recovered from (A6):
$$\tH_n({\bf x},{\bf y},t)=\sum_{j=-\infty}^\infty({\bf H}_{j,n}({\bf y})+{\bf G}_{j,n}({\bf x},{\bf y}))e^{ij\omega t}.$$

After modification of the spatial gradient \rf{grad}, expansion
and separation of mean and fluctuating parts of each Fourier harmonics,
the solenoidality condition \rf{Hsolen} reduces to
$$\nabla_{\bf y}\cdot{\bf H}_{j,n}=0,\eqn{A7}$$
$$\nabla_{\bf x}\cdot{\bf G}_{j,n}+\nabla_{\bf y}\cdot{\bf G}_{j,n-1}=0$$
for all $j$ and $n\ge0$, where it is understood ${\bf G}_{j,n}\equiv0$ for $n<0$.
Here and in what follows the subscripts $\bf x$ and $\bf y$ refer
to differential operators in fast and slow variables, respectively.

Substitution of (A6), \rf{grate} and \rf{grad} into (A5) yields
$$\left.\sum_{n=0}^\infty\right[-ij\omega{\bf H}_{j,n}+{\cal L}_j{\bf G}_{\cdot,n}+\eta\left(
2(\nabla_{\bf x}\cdot\nabla_{\bf y}){\bf G}_{j,n-1}
+\nabla^2_{\bf y}({\bf H}_{j,n-2}+{\bf G}_{j,n-2})\right)$$
$$+\nabla_{\bf x}\times\left({\bf W}\times{\bf H}_{j-1,n}+{\bf U}\times{\bf H}_{j,n}+\overline{\bf W}\times{\bf H}_{j+1,n}\right)$$
$$+\nabla_{\bf y}\times\left({\bf W}\times({\bf H}_{j-1,n-1}+{\bf G}_{j-1,n-1})
+{\bf U}\times({\bf H}_{j,n-1}+{\bf G}_{j,n-1})\right.$$
$$\left.+\overline{\bf W}\times({\bf H}_{j+1,n-1}+{\bf G}_{j+1,n-1})\right)
\left.-\sum_{m=0}^n\lambda_{n-m}({\bf H}_{j,m}+{\bf G}_{j,m})\right]\epsilon^n=0.\eqn{A8}$$
Here it is denoted
$${\cal L}_j{\bf F}\equiv\eta\nabla^2_{\bf x}{\bf F}_j
+\nabla_{\bf x}\times\left({\bf W}\times{\bf F}_{j-1}+{\bf U}\times{\bf F}_j
+\overline{\bf W}\times{\bf F}_{j+1}\right)-ij\omega{\bf F}_j,$$
where ${\bf F}_j$ are Fourier components of
$${\bf F}({\bf x},{\bf y},t)=\sum_{j=-\infty}^\infty{\bf F}_j({\bf x},{\bf y})e^{ij\omega t}.$$

We make a final technical assumption that for any function ${\bf f}({\bf x},t)$,\break
$T$-periodic in time and $2\pi$-periodic in space,
such that $\la{\bf f}\ra=0$ for all $t$, the problem ${\cal L}{\bf F}=\bf f$
has a unique small-scale solenoidal solution with a vanishing spatial mean,
which has the same time and space periodicities, as the flow.
(Equivalently, the magnetic induction operator $\cal L$ is assumed to have
a trivial kernel.) Generically this condition holds.

We proceed by successively equating the mean and the fluctuating part
of each term of the series (A8) to zero.

\mi
$i$. The leading $(n=0)$ term of (A8) takes the form
$${\cal L}_j{\bf G}_{\cdot,0}+({\bf H}_{j-1,0}\cdot\nabla_{\bf x}){\bf W}
+({\bf H}_{j,0}\cdot\nabla_{\bf x}){\bf U}
+({\bf H}_{j+1,0}\cdot\nabla_{\bf x})\overline{\bf W}$$
$$=\lambda_0({\bf H}_{j,0}+{\bf G}_{j,0})+ij\omega{\bf H}_{j,0}.\eqn{A9}$$
The mean of (A9) is
$$0=(ij\omega+\lambda_0){\bf H}_{j,0}.\eqn{A10}$$
Thus it can be assumed
$${\bf H}_{j,0}=0\quad\forall j\ne0;\qquad\lambda_0=0\eqn{A11}$$
(this is a normalisation condition: any other formal solution to (A10):\break
${\bf H}_{j,0}=0\ \forall j\ne J,\ \lambda_0=-iJ\omega$ for $J\ne0$,
does not represent any new eigensolution to the original Floquet problem,
since solutions $\bf H$ to the Floquet problem \rf{floquet} are defined up to
a factor $Ce^{iJ\omega t}$).

Consequently, the fluctuating part of (A9) yields by linearity
$${\bf G}_{j,0}=\sum_{k=1}^3{\bf S}_{j,k}({\bf x}){\bf H}^k_{0,0}({\bf y}),\eqn{A12}$$
where vector fields ${\bf S}_{j,k}$ satisfy
$${\cal L}_j{\bf S}_{\cdot,k}=-{\partial\over\partial x_k}\left(
\delta^j_1{\bf W}+\delta^j_0{\bf U}+\delta^j_{-1}\overline{\bf W}\right),\eqn{A13}$$
which is a representation of the first auxiliary problem \rf{Seq} involving
notation (A2). The problem (A13) has a unique
solution by the assumption that the kernel of $\cal L$ is empty. It is evident
from (A13) and from the definition of the operators ${\cal L}_j$ that
$${\bf S}_{-j,k}=\overline{\bf S}_{j,k};\quad\nabla_{\bf x}\cdot{\bf S}_{j,k}=0\quad\forall j,k.$$
Parity invariance of the flow \rf{parity} implies that
parity anti-invariant vector fields are an invariant subspace of $\cal L$,
and since the right-hand side of (A13) is parity anti-invariant, so are ${\bf S}_{j,k}$:
$${\bf S}_{j,k}({\bf x})={\bf S}_{j,k}(-{\bf x}).\eqn{A14}$$
Divergence (in fast variables) of (A13) implies that ${\bf S}_{j,k}$ are solenoidal.

\mi
$ii$. The second $(n=1)$ term of (A8) reduces with the use of (A3), (A7) and (A11) to
$${\cal L}_j{\bf G}_{\cdot,1}+2\eta(\nabla_{\bf x}\cdot\nabla_{\bf y}){\bf G}_{j,0}
+({\bf H}_{j-1,1}\cdot\nabla_{\bf x}){\bf W}
+({\bf H}_{j,1}\cdot\nabla_{\bf x}){\bf U}
+({\bf H}_{j+1,1}\cdot\nabla_{\bf x})\overline{\bf W}$$
$$+\nabla_{\bf y}\times\left({\bf W}\times{\bf G}_{j-1,0}
+{\bf U}\times{\bf G}_{j,0}+\overline{\bf W}\times{\bf G}_{j+1,0}\right)$$
$$-\left((\delta^j_1{\bf W}+\delta^j_0{\bf U}+
\delta^j_{-1}\overline{\bf W})\cdot\nabla_{\bf y}\right){\bf H}_{0,0}
=\lambda_1(\delta^j_0{\bf H}_{0,0}+{\bf G}_{j,0})+ij\omega{\bf H}_{j,1}.\eqn{A15}$$
Upon substitution of (A12) the mean of (A15) becomes
$$\nabla_{\bf y}\times\sum_{k=1}^3\la{\bf W}\times{\bf S}_{j-1,k}+{\bf U}\times{\bf S}_{j,k}
+\overline{\bf W}\times{\bf S}_{j+1,k}\ra
{\bf H}^k_{0,0}=\lambda_1\delta^j_0{\bf H}_{0,0}+ij\omega{\bf H}_{j,1}.$$
In view of (A4) and (A14), the averaged cross products at the left-hand side
of this equation vanish and thus
$${\bf H}_{j,1}=0\quad\forall j\ne0;\qquad\lambda_1=0.\eqn{A16}$$

After (A16) and the representations (A12) are plugged in, the fluctuating
part of (A15) becomes
\pagebreak
$${\cal L}_j{\bf G}_{\cdot,1}=-({\bf H}_{0,1}\cdot\nabla_{\bf x})
\left(\delta^j_1{\bf W}+\delta^j_0{\bf U}+\delta^j_{-1}\overline{\bf W}\right)$$
$$+\left.\sum_{k=1}^3\sum_{m=1}^3\right[-2\eta{\partial{\bf S}_{j,k}\over\partial x_m}
+{\bf W}^m\left({\bf S}_{j-1,k}+\delta^j_1{\bf e}_k\right)-{\bf W}{\bf S}^m_{j-1,k}\eqn{A17}$$
$$+{\bf U}^m\left({\bf S}_{j,k}+\delta^j_0{\bf e}_k\right)-{\bf U}{\bf S}^m_{j,k}
+\overline{\bf W}^m\left({\bf S}_{j+1,k}+\delta^j_{-1}{\bf e}_k\right)
-\overline{\bf W}{\bf S}^m_{j+1,k}\left]{\partial{\bf H}^k_{0,0}\over\partial y_m}\right..$$
Hence by linearity
$${\bf G}_{j,1}=\sum_{k=1}^3{\bf S}_{j,k}({\bf x}){\bf H}^k_{0,1}({\bf y})+
\sum_{k=1}^3\sum_{m=1}^3\Gmm_{j,m,k}({\bf x})
{\partial{\bf H}^k_{0,0}\over\partial y_m}({\bf y}),\eqn{A18}$$
where vector fields $\Gmm_{j,m,k}({\bf x})$ satisfy
$${\cal L}_j\Gmm_{\cdot,m,k}=-2\eta{\partial{\bf S}_{j,k}\over\partial x_m}
+{\bf W}^m\left({\bf S}_{j-1,k}+\delta^j_1{\bf e}_k\right)-{\bf W}{\bf S}^m_{j-1,k}$$
$$+{\bf U}^m\left({\bf S}_{j,k}+\delta^j_0{\bf e}_k\right)-{\bf U}{\bf S}^m_{j,k}
+\overline{\bf W}^m\left({\bf S}_{j+1,k}+\delta^j_{-1}{\bf e}_k\right)
-\overline{\bf W}{\bf S}^m_{j+1,k},\eqn{A19}$$
which is a representation of the second auxiliary problem \rf{Geq} involving
notation (A2). By standard arguments it is verified that
$$\Gmm_{-j,m,k}=\overline\Gmm_{j,m,k};\quad
\Gmm_{j,m,k}({\bf x})=-\Gmm_{j,m,k}(-{\bf x}).$$
Divergence (in fast variables) of (A19) implies
$\nabla_{\bf x}\cdot\Gmm_{j,m,k}+{\bf S}^m_{j,k}=0$.

\mi
$iii$. The third $(n=2)$ term of (A8) by virtue of (A11) and (A16) reduces to
$${\cal L}_j{\bf G}_{\cdot,2}+\eta\left(2(\nabla_{\bf x}\cdot\nabla_{\bf y}){\bf G}_{j,1}
+\nabla^2_{\bf y}(\delta^j_0{\bf H}_{0,0}+{\bf G}_{j,0})\right)$$
$$+({\bf H}_{j-1,2}\cdot\nabla_{\bf x}){\bf W}
+({\bf H}_{j,2}\cdot\nabla_{\bf x}){\bf U}
+({\bf H}_{j+1,2}\cdot\nabla_{\bf x})\overline{\bf W}$$
$$+\nabla_{\bf y}\times\left({\bf W}\times{\bf G}_{j-1,1}+{\bf U}\times{\bf G}_{j,1}
+\overline{\bf W}\times{\bf G}_{j+1,1}\right)$$
$$-\left((\delta^j_1{\bf W}+\delta^j_0{\bf U}+\delta^j_{-1}\overline{\bf W})
\cdot\nabla_{\bf y}\right){\bf H}_{0,1}
=\lambda_2(\delta^j_0{\bf H}_{0,0}+{\bf G}_{j,0})+ij\omega{\bf H}_{j,2}.\eqn{A20}$$
In view of (A18), (A4) and (A14) the mean of this equations is
$$\eta\delta^j_0\nabla^2_{\bf y}{\bf H}_{0,0}+\nabla_{\bf y}\times
\sum_{m=1}^3\sum_{k=1}^3\la{\bf W}\times\Gmm_{j-1,m,k}+{\bf U}\times\Gmm_{j,m,k}+
\overline{\bf W}\times\Gmm_{j+1,m,k}\ra{\partial{\bf H}^k_{0,0}\over\partial y_m}$$
$$=\lambda_2\delta^j_0{\bf H}_{0,0}+ij\omega{\bf H}_{j,2}.\eqn{A21}$$
Thus the leading terms of the expansions of the mean magnetic field,
${\bf H}_{0,0}$, and of the growth rate, $\lambda_2$, are a solution to the eigenvalue problem
$${\cal M}{\bf H}_{0,0}\equiv\eta\nabla^2{\bf H}_{0,0}
+\nabla_{\bf y}\times\sum_{m=1}^3\sum_{k=1}^3\left(2{\rm Re}\la\overline{\bf W}\times\Gmm_{1,m,k}\ra
+\la{\bf U}\times\Gmm_{0,m,k}\ra\right){\partial{\bf H}^k_{0,0}\over\partial y_m}$$
$$=\lambda_2{\bf H}_{0,0},\eqn{A22}$$
arising from (A21) for $j=0$. (A22) is equivalent to
\rf{leadingtermseq} in view of (A2). Subsequently one obtains from (A21)
$${\bf H}_{j,2}={1\over ij\omega}\nabla_{\bf y}\times
\sum_{m=1}^3\sum_{k=1}^3\la{\bf W}\times\Gmm_{j-1,m,k}+{\bf U}\times\Gmm_{j,m,k}+\overline{\bf W}\times\Gmm_{j+1,m,k}\ra
{\partial{\bf H}^k_{0,0}\over\partial y_m}$$
for any $j\ne0$. Hence now the quantities ${\bf G}_{j,0}$ are also
entirely determined by (A12).

Bounded solutions to the eigenvalue problem (A22) are Fourier harmonics
${\bf H}_{0,0}=\ht e^{i\bf qy}$, satisfying
the orthogonality condition \rf{ortho} and
$$\eta|{\bf q}|^2\ht+{\bf q}\times\sum_{m=1}^3(2{\rm Re}\la{\bf W}\times\Gmm_{-1,m,k}\ra
+\la{\bf U}\times\Gmm_{0,m,k}\ra){\bf q}_m\ht_k=-\lambda_2\ht,\eqn{A23}$$
which is an equivalent of \rf{oldeigen} in new notation (A2).

The $\bf x$-dependent prefactors in front of unknown vector fields
${\bf H}^k_{0,2}$ and $\partial{\bf H}^k_{0,1}/\partial y_m$
in the fluctuating part of (A20),
$${\cal L}_j{\bf G}_{\cdot,2}=-\eta\left(2\sum_{k=1}^3\sum_{m=1}^3
\left({\partial{\bf S}_{j,k}\over\partial x_m}{\partial{\bf H}^k_{0,1}\over\partial y_m}+
\sum_{l=1}^3{\partial\Gmm_{j,m,k}\over\partial x_l}
{\partial^2{\bf H}^k_{0,0}\over\partial y_m\partial y_l}\right)
+\nabla^2_{\bf y}{\bf G}_{j,0}\right)$$
$$-({\bf H}_{j-1,2}\cdot\nabla_{\bf x}){\bf W}-({\bf H}_{j,2}\cdot\nabla_{\bf x}){\bf U}
-({\bf H}_{j+1,2}\cdot\nabla_{\bf x})\overline{\bf W}$$
$$+\left((\delta^j_1{\bf W}+\delta^j_0{\bf U}+
\delta^j_{-1}\overline{\bf W})\cdot\nabla_{\bf y}\right){\bf H}_{0,1}\eqn{A24}$$
$$-\nabla_{\bf y}\times\left(
\sum_{k=1}^3\left({\bf W}\times{\bf S}_{j-1,k}+{\bf U}\times{\bf S}_{j,k}+
\overline{\bf W}\times{\bf S}_{j+1,k}\right){\bf H}^k_{0,1}\right.$$
$$+\left.\sum_{k=1}^3\sum_{m=1}^3\lb{\bf W}\times\Gmm_{j-1,m,k}+{\bf U}\times\Gmm_{j,m,k}+
\overline{\bf W}\times\Gmm_{j+1,m,k}\rb
{\partial{\bf H}^k_{0,0}\over\partial y_m}\right)+\lambda_2{\bf G}_{j,0},$$
are the same as those in front of ${\bf H}^k_{0,1}$ and
$\partial{\bf H}^k_{0,0}/\partial y_m$, respectively, in (A17), which
by linearity implies a representation
$${\bf G}_{j,2}=\sum_{k=1}^3{\bf S}_{j,k}({\bf x}){\bf H}^k_{0,2}({\bf y})+
\sum_{k=1}^3\sum_{m=1}^3\Gmm_{j,m,k}({\bf x})
{\partial{\bf H}^k_{0,1}\over\partial y_m}({\bf y})+{\bf Q}_{j,2}({\bf x},{\bf y}).$$
Vector fields ${\bf Q}_{j,2}$ can be found from a system equations,
obtained from (A24) by changing ${\bf G}_{\cdot,2}\to{\bf Q}_{\cdot,2}$
and dropping all terms involving ${\bf H}^k_{0,2}$ or derivatives of
${\bf H}^k_{0,1}$ (the right-hand sides of the resultant
equations are at this stage known).

\mi
$iv$. Let $\lambda'_2$ and ${\bf h}'$ be the second eigenvalue
and the associated eigenvector, satisfying (A23) and \rf{ortho}.
Subsequent $(n>2)$ terms of (A8) provide a hierarchy of equations,
which can be solved under the condition $\lambda_2\ne\lambda'_2$.
Equations for $n<N$ yield:

\noindent
$\bullet$ vector fields ${\bf H}_{j,n}$ for all $j\ne0$ and $n<N$;

\noindent
$\bullet$ vector fields ${\bf H}_{0,n}$ for all $n<N-2$;

\noindent
$\bullet$ vector fields ${\bf G}_{j,n}$ for all $j$ and $n<N-2$;

\noindent
$\bullet$ representations of ${\bf G}_{j,n}$ of the form
$${\bf G}_{j,n}=\sum_{k=1}^3{\bf S}_{j,k}({\bf x}){\bf H}^k_{0,n}({\bf y})+
\sum_{k=1}^3\sum_{m=1}^3\Gmm_{j,m,k}({\bf x})
{\partial{\bf H}^k_{0,n-1}\over\partial y_m}({\bf y})+{\bf Q}_{j,n}({\bf x},{\bf y})\eqn{A25}$$

for $n=N-1$ and $n=N-2$ with known vector fields
${\bf Q}_{j,n}$, $\la{\bf Q}_{j,n}\ra=0$;

\noindent
$\bullet$ quantities $\lambda_n$ for all $n<N$.

Upon substitution of (A25) for $n=N-1$ the mean
of the equation corresponding to $n=N$ in (A8) becomes
$$\eta\nabla^2{\bf H}_{j,N-2}+\nabla_{\bf y}\times\sum_{m=1}^3\sum_{k=1}^3
\la{\bf W}\times\Gmm_{j-1,m,k}+{\bf U}\times\Gmm_{j,m,k}+\overline{\bf W}\times\Gmm_{j+1,m,k}\ra
{\partial{\bf H}^k_{0,N-2}\over\partial y_m}$$
$$+\nabla_{\bf y}\times\la{\bf W}\times{\bf Q}_{j-1,N-1}+{\bf U}\times{\bf Q}_{j,N-1}
+\overline{\bf W}\times{\bf Q}_{j+1,N-1}\ra$$
$$=\sum_{m=0}^{N-2}\lambda_{N-m}{\bf H}_{j,m}+ij\omega{\bf H}_{j,N}.\eqn{A26}$$
Consider (A26) for $j=0$:
$$({\cal M}-\lambda_2){\bf H}_{0,N-2}-\lambda_N{\bf H}_{0,0}$$
$$=\sum_{m=1}^{N-3}\lambda_{N-m}{\bf H}_{0,m}-\nabla_{\bf y}\times\left(
2{\rm Re}\la\overline{\bf W}\times{\bf Q}_{1,N-1}\ra
+\la{\bf U}\times{\bf Q}_{0,N-1}\ra\right),\eqn{A27}$$
where the right-hand side is a known vector field.
Projecting this equation out in the direction of ${\bf H}_{0,0}$
one can uniquely determine $\lambda_N$. In the complementary
invariant subspace of $\cal M$ the operator
${\cal M}-\lambda_2$ is invertible, and thus ${\bf H}_{0,N-2}$
can be determined from (A27) up to an arbitrary multiple of ${\bf H}_{0,0}$,
which we can demand to vanish. Now ${\bf H}_{j,N}$ for $j\ne0$ can be
found from (A26), and ${\bf G}_{j,N-2}$ are determined by (A25) for $n=N-2$.

The fluctuating part of the equation corresponding to $n=N$ in (A8) becomes
after the substitution of (A25) for $n=N-1$
$${\cal L}_j{\bf G}_{\cdot,N}=-\eta\left(2\sum_{k=1}^3\sum_{m=1}^3\left(
{\partial{\bf S}_{j,k}\over\partial x_m}{\partial{\bf H}^k_{0,N-1}\over\partial y_m}
+\sum_{l=1}^3{\partial\Gmm_{j,m,k}\over\partial x_l}
{\partial^2{\bf H}^k_{0,N-2}\over\partial y_l\partial y_m}\right)\right.$$
$$\left.+2(\nabla_{\bf x}\cdot\nabla_{\bf y}){\bf Q}_{j,N-1}
+\nabla^2_{\bf y}{\bf G}_{j,N-2}\right)
-({\bf H}_{j-1,N}\cdot\nabla_{\bf x}){\bf W}+({\bf W}\cdot\nabla_{\bf y}){\bf H}_{j-1,N-1}$$
$$-({\bf H}_{j,N}\cdot\nabla_{\bf x}){\bf U}+({\bf U}\cdot\nabla_{\bf y}){\bf H}_{j,N-1}
-({\bf H}_{j+1,N}\cdot\nabla_{\bf x})\overline{\bf W}
+(\overline{\bf W}\cdot\nabla_{\bf y}){\bf H}_{j+1,N-1}$$
$$-\nabla_{\bf y}\times\left[\sum_{k=1}^3\left({\bf W}\times{\bf S}_{j-1,k}+{\bf U}\times{\bf S}_{j,k}+
\overline{\bf W}\times{\bf S}_{j+1,k}\right){\bf H}^k_{0,N-1}\right.\eqn{A28}$$
$$+\sum_{k=1}^3\sum_{m=1}^3\lb{\bf W}\times\Gmm_{j-1,m,k}+{\bf U}\times\Gmm_{j,m,k}+
\overline{\bf W}\times\Gmm_{j+1,m,k}\rb
{\partial{\bf H}^k_{0,N-2}\over\partial y_m}$$
$$+\lb{\bf W}\times{\bf Q}_{j-1,N-1}+{\bf U}\times{\bf Q}_{j,N-1}+
\overline{\bf W}\times{\bf Q}_{j+1,N-1})\rb\left]
+\sum_{m=0}^{N-2}\lambda_{N-m}{\bf G}_{j,m}\right..$$
Like in the case of (A24), the structure of this equation implies by linearity
the representation (A25) of ${\bf G}_{j,N}$, where the quantities
${\bf Q}_{j,N}$ are uniquely defined by
$${\cal L}_j{\bf Q}_{\cdot,N}=-\eta\left(2\sum_{k=1}^3\sum_{m=1}^3
\sum_{l=1}^3{\partial\Gmm_{j,m,k}\over\partial x_l}
{\partial^2{\bf H}^k_{0,N-2}\over\partial y_l\partial y_m}
+2(\nabla_{\bf x}\cdot\nabla_{\bf y}){\bf Q}_{j,N-1}+\nabla^2_{\bf y}{\bf G}_{j,N-2}\right)$$
$$-(1-\delta^j_1)({\bf H}_{j-1,N}\cdot\nabla_{\bf x}){\bf W}
-(1-\delta^j_0)({\bf H}_{j,N}\cdot\nabla_{\bf x}){\bf U}
-(1-\delta^j_{-1})({\bf H}_{j+1,N}\cdot\nabla_{\bf x})\overline{\bf W}$$
$$+(1-\delta^j_1)({\bf W}\cdot\nabla_{\bf y}){\bf H}_{j-1,N-1}
+(1-\delta^j_1)({\bf U}\cdot\nabla_{\bf y}){\bf H}_{j,N-1}
+(1-\delta^j_{-1})(\overline{\bf W}\cdot\nabla_{\bf y}){\bf H}_{j+1,N-1}$$
$$-\nabla_{\bf y}\times\left[\sum_{k=1}^3\sum_{m=1}^3
\lb{\bf W}\times\Gmm_{j-1,m,k}+{\bf U}\times\Gmm_{j,m,k}+\overline{\bf W}\times\Gmm_{j+1,m,k}\rb
{\partial{\bf H}^k_{0,N-2}\over\partial y_m}\right.$$
$$+\lb{\bf W}\times{\bf Q}_{j-1,N-1}+{\bf U}\times{\bf Q}_{j,N-1}+
\overline{\bf W}\times{\bf Q}_{j+1,N-1})\rb\left]
+\sum_{m=0}^{N-2}\lambda_{N-m}{\bf G}_{j,m}\right.$$
(the right-hand side of this equation is known). This equation was
obtained by omitting in (A28) all terms, involving
${\bf H}_{0,N}$ or derivatives of ${\bf H}_{0,N-1}$, and changing
${\bf G}_{\cdot,N}$ to ${\bf Q}_{\cdot,N}$.

Thus a complete asymptotic expansion of magnetic modes and their growth rates
is constructed. Like in the case of a stationary velocity, it can be easily
verified that
$${\bf H}_{0,0}={\bf h}e^{i\bf qy},\quad
{\bf H}_{0,n}=\chi_n{\bf h}'e^{i\bf qy}\quad\forall n>0,\quad
{\bf H}_{j,n}={\bf h}_{j,n}e^{i\bf qy}\quad\forall n,\ j\ne0$$
(where ${\bf h}_{j,n}=$const, ${\bf h}_{j,n}\cdot{\bf q}=0$),
$${\bf G}_{j,n}={\bf g}_n({\bf x})e^{i\bf qy}\quad\forall n\ge0$$
and thus the eigenmode admits a representation
$${\bf H}=e^{i\epsilon\bf qx}{\bf h}({\bf x},t),\quad{\bf q}=\hbox{const}.\eqn{A29}$$
(For this reason for constructions of section $iv$
it was sufficient to demand that $\lambda'\ne\lambda_2$, and not that
${\cal M}-\lambda_2$ is invertible in the whole domain.)
This stems from the fact that for the velocity (A1)
the domain of the magnetic induction operator splits into invariant subspaces, each comprised of
vector fields (A29) and categorised by wavevectors $\bf q$.

Analyzing parity of solutions of the hierarchy of equations (A8) constructed in this Appendix,
one finds that all ${\bf G}_{j,n}$ with even indices $n$ are parity anti-invariant;
all ${\bf G}_{j,n}$ with odd indices $n$ are parity-invariant;
$\la{\bf H}_{j,n}\ra=\bf 0$ for any odd $n$; and $\lambda_n=0$ for any odd $n$.

\end{document}